\documentclass[a4paper,fleqn]{cas-sc}

\usepackage[numbers,sort&compress]{natbib}

\usepackage{graphicx}%
\usepackage{multirow}%
\usepackage{amsmath,amssymb,amsfonts}%
\usepackage{amsthm}%
\usepackage{mathrsfs}%
\usepackage[title]{appendix}%
\usepackage{xcolor}%
\usepackage{textcomp}%
\usepackage{manyfoot}%
\usepackage{booktabs}%
\usepackage{quantikz}%
\usepackage{algorithm}%
\usepackage{algorithmicx}%
\usepackage{algpseudocode}%
\usepackage{listings}%
\usepackage{braket}
\usepackage{bm}
\usepackage{subcaption}

\usepackage{placeins}

\def\tsc#1{\csdef{#1}{\textsc{\lowercase{#1}}\xspace}}
\tsc{WGM}
\tsc{QE}
\tsc{EP}
\tsc{PMS}
\tsc{BEC}
\tsc{DE}

\begin{document}
\let\WriteBookmarks\relax
\shorttitle{Hybrid Quantum Image Preparation via JPEG Compression}
\shortauthors{E Rezaei Fard Boosari}

\title [mode = title]{Hybrid Quantum Image Preparation via JPEG Compression}

\author{Emad Rezaei Fard Boosari}[type=editor,
                        auid=000,bioid=1,                                               
                        orcid=0000-0002-4774-3915]
\cormark[1]
\fnmark[1]
\ead{emad.boosari@gmail.com}

\credit{Conceptualization of this study, Methodology, Software, Data curation, Writing - Original draft preparation}


\cortext[cor1]{Corresponding author}

\begin{abstract}
	We present a hybrid classical--quantum image preparation scheme that reduces the quantum implementation cost of image loading for quantum pixel information encoding (QPIE).
	The proposed method, termed \emph{JPEG-assisted QPIE} (JQPIE), loads only the quantized JPEG coefficients into a quantum register, leading to substantial reductions in \texttt{CX} gate count and circuit depth while preserving reconstruction quality comparable to classical JPEG compression.
	We develop two variants of the hybrid strategy.
	The first realizes the complete JPEG decompression pipeline coherently by implementing inverse quantization via a block-encoded unitary operator.
	The second, referred to as \emph{quantization-free JQPIE} (QF-JQPIE), omits quantization altogether, thereby avoiding the probabilistic nature of block-encoded quantization.
	Numerical simulations on standard benchmark image datasets (USC--SIPI and Kodak) demonstrate that both variants achieve significant constant-factor reductions in \texttt{CX} gate count and circuit depth relative to direct QPIE loading, while maintaining high reconstruction quality as measured by PSNR and SSIM.
\end{abstract}

\begin{keywords}
	Quantum image processing \sep
	Amplitude encoding \sep
	JPEG compression \sep
	QSP \sep
	Hybrid quantum--classical algorithms
\end{keywords}

\maketitle

\section{Introduction}\label{sec:introduction}

Quantum image processing (QIMP) has emerged as a promising approach to accelerate digital image processing tasks across several domains, including artificial intelligence~\cite{goodfellow2016deep} and computer vision~\cite{krizhevsky2012imagenet,szeliski2022computer}. Despite substantial progress in QIMP~\cite{chetia2021quantum,el2016quantum,liu2023quantum,wang2015quantum,zhou2017quantum, li2024quantum}, efficiently loading images onto a quantum register remains a major bottleneck, as encoding pixel intensities into quantum amplitudes can incur an exponential gate cost on quantum hardware. This work addresses this challenge by proposing a quantum state preparation (QSP) scheme that encodes classical image pixels into the amplitudes of quantum states in a hybrid manner.

Recently, a simple hybrid QSP (HQSP) framework has been introduced to load compressible data into a quantum register with reduced gate complexity and circuit depth~\cite{boosari2025hybrid}.
While the method has primarily been applied to one-dimensional signals, the framework is generally applicable to other types of compressible data, such as images.
HQSP relies on a reversible classical compression strategy together with a corresponding quantum procedure that coherently implements data decompression within a quantum circuit.
For image data, a natural candidate satisfying these requirements is the Joint Photographic Experts Group (JPEG) compression standard~\cite{pennebaker1992jpeg}, which is based on the energy compaction property of the block-wise discrete cosine transform (DCT).
Notably, a quantum implementation of the fast DCT for fixed $8\times8$ blocks is available and requires only a shallow circuit with a small, constant number of two-qubit gates~\cite{tang2019quantum}.

The JPEG algorithm has been already deployed in hybrid regimes \cite{jiang2018novel, haque2023advanced, haque2024block} to reduce the quantum preparation cost. But all of them focus on pure discrete quantum image representation (QIMRs) \cite{zhang2013neqr, li2018quantum2, jiang2015quantum, jiang2015quantum2, nasr2021efficient} which store pixel values in the computational basis and adopt non-unitary operation such as element-wise multiplication in quantum circuit. 
Despite this flexibility, computational basis embedding typically require a large number of ancillary qubits for quantum matrix multiplication, resulting in wide circuits and limiting the overall quantum resource reduction in practice.

By contrast, mixed quantum image representations (QIMRs)~\cite{le2011flexible,sun2023,khan2019improved,sun2011multi,yao2017quantum}, which encode pixel intensities in continuous quantum amplitudes while using the discrete computational basis to represent spatial positions, have remained largely unexplored in the context of JPEG compression.
The primary reason for this limited adoption is that mixed QIMRs cannot directly implement inherently non-unitary operations such as quantization within a coherent quantum circuit. 
However a quantum version of JPEG algorithm has been introduced ~\cite{roncallo2023quantum} based on the JPEG algorithm intuitions, namely the suppression of high-frequency components but relies on quantum Fourier transformation and down-sampling rather than implementing the standard JPEG compression workflow.

In this paper, we propose a hybrid classical–quantum image preparation scheme, termed JQPIE, which combines the classic JPEG compression algorithm with the QPIE amplitude-encoding representation~\cite{yao2017quantum}. 
In QPIE, normalized pixel intensities are encoded directly into quantum amplitudes by using exponentially many \texttt{CX} and rotation gates in the number of pixels, while JQPIE framework reduces this cost by loading only the quantized JPEG coefficients into the quantum register and reconstructing the full image through a coherent quantum decompression procedure.
A central challenge in this approach is the implementation of JPEG quantization operation, which is inherently lossy and non-unitary.
To overcome this limitation, we realize the inverse quantization step via a block-encoded diagonal operator~\cite{gilyen2019quantum}, enabling a fully coherent approximation of the JPEG decompression pipeline within a unitary quantum circuit.

Beyond the JQPIE construction, which introduces probabilistic overhead through block-encoded operations, we also propose a quantization-free variant, referred to as \emph{QF-JQPIE}.
This alternative approach exploits the intrinsic energy compaction of the block-wise DCT, together with truncation of zigzag-reordered coefficients, to suppress high-frequency components while retaining the dominant image content within a reduced Hilbert space.
By omitting JPEG quantization entirely, QF-JQPIE eliminates the need for ancillary qubits, block encoding, amplitude amplification, and post-selection, resulting in a fully unitary image preparation circuit.

Both hybrid algorithms are evaluated using numerical simulations on standard benchmark images from the USC--SIPI and Kodak datasets.
Reconstruction quality is assessed using peak signal-to-noise ratio (PSNR)~\cite{gonzalez2009digital} and structural similarity index (SSIM)~\cite{wang2004image}, while quantum resource requirements are analyzed in terms of \texttt{CX} gate counts and circuit depth, with comparisons against the original QPIE framework.
All quantum circuits are constructed and simulated using standard quantum software frameworks, including \texttt{Qiskit} \cite{javadi2024quantum} and \texttt{Qibo} \cite{qibo_paper}, enabling a faithful assessment of circuit-level resource requirements.

The remainder of this paper is organized as follows.
Section~\ref{sec:preliminaries} introduces the essential background required for the proposed framework.
Section~\ref{sec:hybrid_QIP} presents the JQPIE method and its quantization-free variant.
Section~\ref{sec:results_discussion} reports numerical results and discusses their implications for reconstruction fidelity and quantum resource efficiency.
Finally, Section~\ref{sec:conclusion} concludes the main findings and outlines directions for future work.

\section{Preliminaries}\label{sec:preliminaries}

This section introduces the definitions, notation, and background concepts required for the construction of the proposed hybrid frameworks.
We briefly review the QPIE representation~\cite{yao2017quantum} and the HQSP paradigm for compressible data~\cite{boosari2025hybrid}.
We then review the relevant components of the classical JPEG compression algorithm~\cite{pennebaker1992jpeg} that are used throughout this work.
Finally, we define the image-quality metrics employed to evaluate reconstruction quality.

\subsection{Quantum Pixel Information Encoding}

One of the most direct approaches for embedding classical images into a quantum register is the QPIE scheme. In QPIE, pixel intensities are mapped to the amplitudes of a quantum state, while pixel coordinates are encoded in the computational basis~\cite{yao2017quantum}.

Let $I = \left[X(i,j)\right]_{H\times W}$ denote a grayscale image with height $H = 2^h$ and width $W = 2^w$. 
Vectorizing the image in column-major order gives
\begin{equation}
\big[X(0,0), \dots, X(H-1,0), X(0,1), \dots, X(H-1,1),\cdots , X(0,W-1), \dots, X(H-1,W-1)\big]^T,
\end{equation}
which induces the amplitude-encoded quantum image
\begin{equation}\label{eq:QPIE-2D}
	\ket{I} = \frac{1}{\sqrt{\sum_{x,y} X(x,y)^2}} \sum_{x=0}^{H-1}\sum_{y=0}^{W-1} X(x,y)\ket{x}\ket{y},
\end{equation}
where, for the sake of simplicity, we use the shorthand $\ket{\alpha}\otimes\ket{\beta}\otimes\cdots \otimes\ket{\gamma}\equiv \ket{\alpha}\ket{\beta}\cdots \ket{\gamma} $ throughout this paper.

Although conceptually simple, preparing $\ket{I}$ requires $\mathcal{O}(2^{h+w})$ \texttt{CX} gates and circuit depth~\cite{iten2016, mottonen2004transformation}, which is infeasible for near term devices. This motivates the use of compressed intermediate representations, leading to the hybrid preparation scheme described in the next section.

\subsection{Hybrid QSP}
\label{sec:hybrid_QSP}

The HQSP framework provides an indirect yet resource-efficient method for loading compressible classical data into quantum states ~\cite{boosari2025hybrid}. 
The main idea is to apply classical compression prior to QSP, thereby reducing the number of non-zero coefficients that must be explicitly encoded. 
If the data admits a sparse representation after compression, the required quantum gates scale only polynomially in the number of qubits.

\subsubsection{Classical Compression}

Given a classical vector $\mathbf{x}\in\mathbb{R}^N$, HQSP first performs a reversible classical transform
\begin{equation}
	\mathbf{X} = \mathcal{U}_C \mathbf{x},
\end{equation}
where $\mathcal{U}_C$ concentrates the signal energy into a small set of coefficients.  
If $\mathbf{X}$ is already $d$-sparse with $d\ll N$, it is perfectly reconstructable.  
If not, sparsity is enforced using quantum-compatible coefficient-reduction techniques.

\subsubsection{QSP}
Once a sparse vector is obtained, after normalization one can prepare it by using any sparse QSP (SQSP) algorithm \cite{farias2025quantum, li2024nearly, gonzales2025efficient}. A sparse state encoder $\mathcal{P}$ can be written as
\begin{equation}
	\ket{\phi} = \mathcal{P}\ket{0}^{\otimes n} = \sum_{k \in \mathcal{S}} X^r_k \ket{k},  \qquad |\mathcal{S}| = d,
\end{equation}
where $X_k^r$ is the index $k$ of sparse-normalized vector. The operator $\mathcal{P}$ can be implemented in quantum circuit just by using $\mathcal{O}(\mathrm{poly}(n))$ quantum gates, in contrast to the $\mathcal{O}(2^n)$ cost of exact amplitude encoding algorithms.

\subsubsection{Quantum Decompression}

Within the HQSP framework, decompression is performed coherently by implementing a reversible classical inverse transform as a unitary quantum operation.
Specifically, the quantum state is recovered as
\begin{equation}
	\ket{\Phi} = \mathcal{U}_Q^{-1}\ket{\phi},
\end{equation}
where $\mathcal{U}_Q^{-1}$ is a unitary operator that realizes the classical inverse transformation $\mathcal{U}_C^{-1}$ within a quantum circuit.
If the gate complexity of $\mathcal{U}_Q^{-1}$ scales polynomially with the system size, i.e., $\mathcal{O}(\mathrm{poly}(n))$, the overall gate complexity of the HQSP procedure remains polynomial.

In the present work, we specialize this general HQSP principle to image data by choosing the DCT as the classical transform $\mathcal{U}_C$.
Classical JPEG compression is employed prior to state preparation to enhance sparsity in the DCT domain, thereby reducing the quantum resources required for loading the image.

\subsection{JPEG Algorithm}\label{sec:jpeg}

This section reviews the JPEG compression algorithm, focusing on the components relevant to the proposed hybrid quantum image preparation framework.
The JPEG codec consists of two main stages: \emph{compression} and \emph{decompression}~\cite{pennebaker1992jpeg, gonzalez2018digital}.
In the compression stage, the image is partitioned into blocks, transformed into the frequency domain, quantized, reordered, and entropy encoded.
The decompression stage reverses these operations to reconstruct an approximation of the original image.
The steps outlined in this section provide the classical pre-processing and post-processing structure used throughout this work.

\subsubsection{Compression Process}\label{sec:jpeg_compression}
\begin{figure}[t]
	\centering
	\begin{subfigure}[t]{0.38\linewidth}
		\centering
		\includegraphics[width=\linewidth]{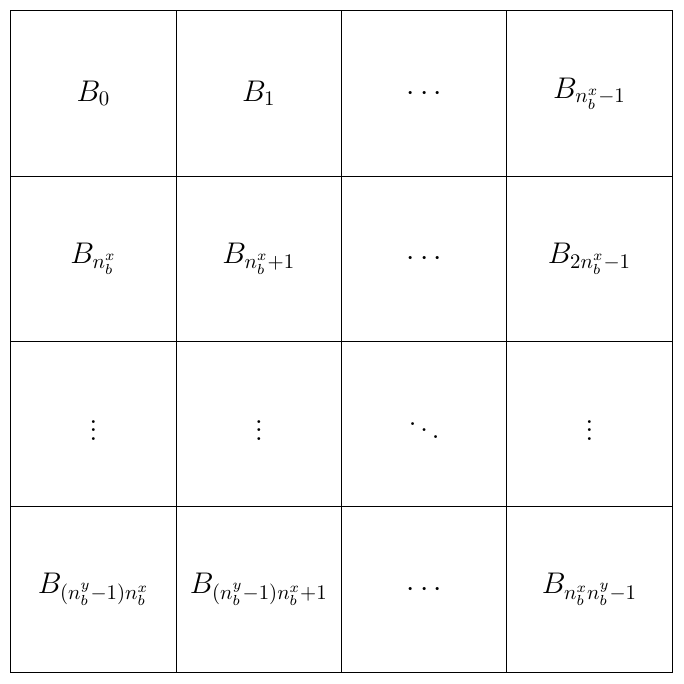}
		\caption{}
		\label{fig:block}
	\end{subfigure}	
	\begin{subfigure}[t]{0.4\linewidth}
		\centering
		\includegraphics[width=\linewidth]{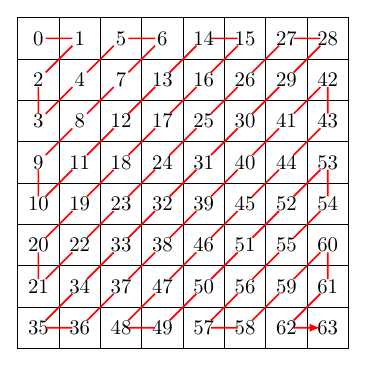}
		\caption{}
		\label{fig:zigzag_reorder}
	\end{subfigure}
	\caption{Classical pre-processing steps used in the hybrid JPEG pipeline.
		(a) Partitioning of an input image into non-overlapping $8\times8$ blocks $B_j$ for localized DCT processing; if the image dimensions are not multiples of eight, zero-padding is applied along the bottom and right edges.
		(b) Zigzag traversal pattern $\pi(k)$ defining the one-dimensional ordering of DCT coefficients within each $8\times8$ block, mapping frequency coordinates $(u,v)$ to sequence indices $k=8u+v$, with low-frequency components appearing first. The ordering begins with low-frequency coefficients (e.g., $\pi(0)=0,\pi(1)=1,\pi(2)=8,\pi(3)=16,\pi(4)=9,\ldots$), which dominate perceptual energy.}
	\label{fig:block_zigzag}
\end{figure}

The JPEG method compresses an image through block-wise operations. 
The input image $I\in\mathbb{R}^{H\times W}$ is partitioned into non-overlapping $8\times8$ blocks, where the number of blocks along each dimension is
\begin{equation}\label{eq:8x8_blocking}
	n_b^x = \left\lceil \frac{H}{8} \right\rceil, \qquad n_b^y = \left\lceil \frac{W}{8} \right\rceil .
\end{equation}
If required, zero-padding is applied along the bottom and right edges to ensure uniform block size.
Each $8\times8$ block $B_j$ contains 64 pixel values,
\begin{equation}\label{eq:single_block_Bj}
	B_j =
	\begin{bmatrix}
		X_j(0,0) & X_j(0,1) & \dots & X_j(0,7) \\
		X_j(1,0) & X_j(1,1) & \dots & X_j(1,7) \\
		\vdots & \vdots & \ddots & \vdots \\
		X_j(7,0) & X_j(7,1) & \dots & X_j(7,7)
	\end{bmatrix},
	\qquad
	j = 0,1,\dots,(n_b^x n_b^y - 1),
\end{equation}
and is processed independently. 
Figure~\ref{fig:block} illustrates an image partitioned into the $8\times8$ non-overlapping blocks.

\paragraph{\textbf{2D DCT.}}
Following the JPEG standard, each $8\times8$ image block $B_j$ is transformed into the frequency domain using the two-dimensional DCT, which is well-known for its strong energy compaction properties~\cite{ahmed2006discrete, pennebaker1992jpeg}.
The 2D DCT of block $B_j$ is defined as
\begin{equation}\label{eq:C_j_equation}
	C_j(u,v) =
	\alpha(u)\,\alpha(v)
	\sum_{x=0}^{7}\sum_{y=0}^{7}
	X_j(x,y)
	\cos\!\left[\frac{(2x+1)u\pi}{16}\right]
	\cos\!\left[\frac{(2y+1)v\pi}{16}\right],
\end{equation}
where $0 \le u,v \le 7$, and the normalization factors are given by
\begin{equation}
	\alpha(k)=
	\begin{cases}
		\sqrt{\tfrac{1}{8}}, & k=0,\\[4pt]
		\sqrt{\tfrac{2}{8}}, & 1 \le k \le 7.
	\end{cases}
\end{equation}

The coefficient $C_j(0,0)$ corresponds to the DC component and captures the average intensity of the block, while the remaining coefficients $C_j(u,v)$ with $(u,v)\neq(0,0)$ represent increasing horizontal and vertical spatial frequencies.
Due to the decorrelating and energy-compaction properties of the DCT, most of the visually significant information is concentrated in the low-frequency coefficients.

Each transformed block can be written in matrix form as
\begin{equation}\label{eq:Cj_block}
	C_j =	
	\begin{bmatrix}
		C_j(0,0) & C_j(0,1) & \cdots & C_j(0,7) \\
		C_j(1,0) & C_j(1,1) & \cdots & C_j(1,7) \\
		\vdots   & \vdots   & \ddots & \vdots \\
		C_j(7,0) & C_j(7,1) & \cdots & C_j(7,7)
	\end{bmatrix}.
\end{equation}
For later use in the JQPIE framework, the DCT coefficients are represented in a one-dimensional vectorized form using a row-major ordering, given by
\begin{equation}\label{eq:V_j}
	V_j = \mathrm{vec}(C_j)
	= \big[C_j(0,0),\, C_j(0,1),\, \ldots,\, C_j(u,v),\, \ldots,\, C_j(7,7)\big]^T
	= \big[C_j(0),\, C_j(1),\, \ldots,\, C_j(k),\, \ldots,\, C_j(63)\big]^T,
\end{equation}
where the one-dimensional index is defined as $k = 8u + v$.

\paragraph{\textbf{Quantization.}}
Quantization exploits the limited sensitivity of the human visual system to high-frequency intensity variations, enabling aggressive suppression of visually insignificant components while preserving perceptual image quality.
Accordingly, after the block-wise DCT, each coefficient block $C_j$ is mapped to a discrete and typically sparse representation through quantization,
\begin{equation}\label{eq:quantization_hat_C}
	\hat{C}_j(u,v)
	=
	\mathrm{round}\!\left(
	\frac{C_j(u,v)}{Q(u,v)}
	\right),
	\qquad 0 \le u,v \le 7 ,
\end{equation}
where $Q(u,v)$ denotes the $(u,v)$-th entry of the standard luminance quantization matrix
\begin{equation}\label{eq:quantization_matrix}
	Q = S .
	\begin{bmatrix}
		16 & 11 & 10 & 16 & 24 & 40 & 51 & 61 \\
		12 & 12 & 14 & 19 & 26 & 58 & 60 & 55 \\
		14 & 13 & 16 & 24 & 40 & 57 & 69 & 56 \\
		14 & 17 & 22 & 29 & 51 & 87 & 80 & 62 \\
		18 & 22 & 37 & 56 & 68 & 109 & 103 & 77 \\
		24 & 35 & 55 & 64 & 81 & 104 & 113 & 92 \\
		49 & 64 & 78 & 87 & 103 & 121 & 120 & 101 \\
		72 & 92 & 95 & 98 & 112 & 100 & 103 & 99
	\end{bmatrix}.
\end{equation}
The scalar parameter $S>0$ controls the overall strength of quantization.
Larger values of $S$ increase the suppression of high-frequency coefficients, leading to stronger sparsity at the expense of reduced reconstruction fidelity, while $S=1$ corresponds to the standard JPEG setting.

Quantization thus performs a frequency-dependent rescaling followed by discretization, preferentially attenuating high-frequency components while preserving dominant low-frequency structure.
As a consequence, each quantized block
\begin{equation}\label{eq:Cj_hat}
	\hat{C}_j =
	\begin{bmatrix}
		\hat{C}_j(0,0) & \hat{C}_j(0,1) & \cdots & \hat{C}_j(0,7) \\
		\hat{C}_j(1,0) & \hat{C}_j(1,1) & \cdots & \hat{C}_j(1,7) \\
		\vdots & \vdots & \ddots & \vdots \\
		\hat{C}_j(7,0) & \hat{C}_j(7,1) & \cdots & \hat{C}_j(7,7)
	\end{bmatrix},
\end{equation}
is typically sparse, containing only a small number of non-zero low-frequency coefficients.
This sparsity is a key structural feature exploited by the HQSP framework and underpins the efficiency gains of the proposed JQPIE method.

For compatibility with subsequent quantum operations, the quantized coefficients are arranged into a one-dimensional representation using vectorization,
\begin{equation}\label{eq:V_j_hat}
	\hat{V}_j = \mathrm{vec}(\hat{C}_j)
	=
	\big[
	\hat{C}_j(0),\, \hat{C}_j(1),\, \ldots,\, \hat{C}_j(k),\, \ldots,\, \hat{C}_j(63)
	\big]^T,
	\qquad k = 8u + v .
\end{equation}

\paragraph{\textbf{Zigzag Transform.}}
After quantization, each block $\hat{C}_j$ is typically sparse, with non-zero coefficients concentrated in the low-frequency region.
In the classical JPEG, these coefficients are reordered into a one-dimensional sequence via the zigzag scan in order to group low-frequency components at the beginning of the stream, thereby improving the efficiency of subsequent entropy coding~\cite{pennebaker1992jpeg}.

The zigzag transform preserves coefficient magnitudes and alters only their ordering. Thus it allows the transformation to be expressed as a linear permutation, which admits a natural unitary realization in a quantum circuit
\begin{equation}\label{eq:Z_j_hat}
	\hat{Z}_j = P\,\hat{V}_j
	= \big[\hat{C}_j(\pi(0)), \hat{C}_j(\pi(1)), \ldots, \hat{C}_j(\pi(k)), \ldots \hat{C}_j(\pi(63))\big]^T,
\end{equation}
where $\pi(k)$ denotes the fixed zigzag traversal pattern illustrated in Fig.~\ref{fig:zigzag_reorder}.
In particular, the zigzag transform enables truncation or selective encoding of the leading coefficients, a feature that is subsequently utilized by the JQPIE framework.

In standard JPEG, after the zigzag transform, entropy coding techniques such as run-length and Huffman coding are applied to further reduce the data size. However, such variable-length encodings are fundamentally incompatible with fixed-dimensional quantum states and unitary quantum evolution. Accordingly, entropy coding is not considered in the JQPIE framework, which targets coherent QSP in a fixed Hilbert space.

\subsubsection{Decompression Process}\label{sec:jpeg_decompression}

The decompression phase reconstruct an approximation of the original image pixels.  
It restores the frequency-domain coefficients of each block and transforms them back into the spatial domain through three successive steps:  
(1) inverse zigzag reordering, (2) inverse quantization, and (3) inverse 2D DCT.

\begin{enumerate}
	\item \textbf{Inverse Zigzag.}  
	Each zigzag-ordered vector $\hat{Z}_j$ (Eq.~\ref{eq:Z_j_hat}) is first mapped back to its earlier position in frequency domain by applying the inverse permutation:
	\begin{equation}
		\hat{V}_j = P^{\dagger} \hat{Z}_j.
	\end{equation}
	Because the permutation matrix $P$ is unitary ($P^{\dagger} = P^{-1}$), this step exactly restores the original coefficient arrangement.
	Reshaping $\hat{V}_j$ back into an $8\times8$ array yields the quantized DCT block $\hat{C}_j(u,v)$.
	
	\item \textbf{Inverse Quantization.}  
	Each quantized coefficient is rescaled by its corresponding element in the quantization matrix $Q(u,v)$:
	\begin{equation}\label{eq:inverse_quantization}
		C_j(u,v) \approx \hat{C}_j(u,v)\,Q(u,v).
	\end{equation}
	This operation reverses the normalization introduced during quantization, although the rounding in Eq.~(\ref{eq:quantization_hat_C}) prevents exact recovery of the original coefficients.

	\item \textbf{Inverse DCT.}  
	The spatial-domain block is then reconstructed as
	\begin{equation}
		X_j(x,y)= \sum_{u=0}^{7}\sum_{v=0}^{7} 	\alpha(u)\alpha(v)\,C_j(u,v) \cos\!\left[\frac{(2x+1)u\pi}{16}\right]
		\cos\!\left[\frac{(2y+1)v\pi}{16}\right].
	\end{equation}
\end{enumerate}

After processing all blocks, the reconstructed image $\widetilde{I}$ is obtained by reassembling the restored blocks in their original spatial arrangement.
Due to the irreversible nature of quantization, $\widetilde{I}$ generally differs from the original image $I$, as high-frequency components are partially discarded.
Nevertheless, for moderate compression levels, the resulting reconstruction preserves high perceptual fidelity and visually salient features.

For convenience and to avoid ambiguity, Table~\ref{tab:notation_summary} summarizes the principal symbols and notation used throughout the JPEG algorithm.
\begin{table}[t]
	\caption{Summary of notation used in the Sec.~\ref{sec:jpeg}.
		Throughout this work, a hat $(\,\hat{\cdot}\,)$ denotes quantized quantities.}
	\label{tab:notation_summary}	
	\centering
	\begin{tabular}{ll}
		\toprule
		\textbf{Symbol} & \textbf{Description} \\
		\midrule
		$I$ & Input grayscale image \\[2pt]
		$j$ & Index labeling image blocks \\[2pt]
		$B_j$ & $j$-th $8\times8$ image block in the spatial domain \\[2pt]
		$X_j(x,y)$ & Pixel intensity at position $(x,y)$ within block $B_j$ \\[2pt]
		$C_j$ & $j$-th block of 2D-DCT coefficients (frequency domain) \\[2pt]		
		$\hat{C}_j$ & Quantized DCT coefficient block of $B_j$ \\[2pt]
		$Q$ & JPEG luminance quantization matrix \\[2pt]
		$\hat{V}_j$ & Vectorized form of the quantized DCT block $\hat{C}_j$ \\[2pt]
		$V_j$ & Vectorized form of the unquantized DCT block $C_j$ \\[2pt]
		$P$ & Zigzag permutation matrix \\[2pt]
		$\pi(k)$ & Zigzag index map \\[2pt]
		$\hat{Z}_j$ & Zigzag-ordered vector of quantized DCT coefficients \\[2pt]
		$Z_j$ & Zigzag-ordered vector of unquantized DCT coefficients \\[2pt]
		$\widetilde{I}$ & Reconstructed (approximate) image \\
		\bottomrule
	\end{tabular}
\end{table}

\subsection{Image–Quality Metrics}\label{sec:metrics}
To assess the quality of reconstructed images obtained after compression and inverse transformation, we employ two widely used classical image-quality metrics: the PSNR and the SSIM.

The PSNR quantifies pixel-wise distortion between the original image $I$ and its reconstruction $\Tilde I$ and is defined as
\begin{equation}
	\mathrm{PSNR}
	=
	10\log_{10}\!\left(\frac{L^2}{\mathrm{MSE}}\right),
	\qquad
	\mathrm{MSE}
	=
	\frac{1}{HW}\sum_{i,j}\big(I(i,j)-\Tilde I(i,j)\big)^2 ,
\end{equation}
where $L$ denotes the maximum possible pixel intensity value (for 8-bit grayscale images, $L=255$), and $HW$ is the image resolution.

In contrast, SSIM evaluates perceptual similarity by comparing luminance, contrast, and structural information between two images and is given by
\begin{equation}
	\mathrm{SSIM}(I,\tilde{I})
	=
	\frac{(2\mu_I\mu_{\tilde{I}}+c_1)(2\sigma_{I\tilde{I}}+c_2)}
	{(\mu_I^2+\mu_{\tilde{I}}^2+c_1)(\sigma_I^2+\sigma_{\tilde{I}}^2+c_2)} .
\end{equation}
Here, $\mu_I$ and $\mu_{\tilde{I}}$ denote the mean intensities, $\sigma_I^2$ and $\sigma_{\tilde{I}}^2$ the variances, $\sigma_{I\tilde{I}}$ the covariance, and $c_1,c_2$ are small constants ensuring numerical stability. 
Higher PSNR and SSIM values indicate better reconstruction quality. These metrics are used throughout this work to quantitatively evaluate the performance of the proposed JQPIE and QF-JQPIE image preparation schemes.

In principle, the accuracy of approximated quantum states is  characterized by state fidelity or trace distance~\cite{nielsen2010quantum}, however, such measures are not fully representative in the present context. The objective of JQPIE is not to faithfully reproduce a quantum state, but rather to reconstruct classical images whose quality is ultimately assessed through human visual perception. Furthermore, perceptual image-quality metrics such as PSNR and SSIM provide a more meaningful evaluation of reconstruction performance, as they directly capture visually observable distortions in the reconstructed images.

\section{Hybrid Quantum Image Preparation}\label{sec:hybrid_QIP}

This section presents two hybrid quantum image preparation algorithms that reduce the \texttt{CX} gate count and circuit depth required by the QPIE representation. 
Both methods build on the HQSP paradigm introduced in Sec.~\ref{sec:hybrid_QSP}: the input image $I$ is first compressed classically using the JPEG pipeline reviewed in Sec.~\ref{sec:jpeg}, a reduced set of transform coefficients is then loaded into a quantum register, and a coherent quantum decompression circuit is applied to reconstruct an approximate image state $\ket{\widetilde{I}}$.

Before introducing the proposed hybrid methods, we empirically characterize the sparsity induced by JPEG on a large benchmark suite comprising 165 images from the USC--SIPI Image Database together with the Kodak image set. 
The dataset spans four categories, textures, aerial images, miscellaneous images, and Kodak images, and covers a broad range of resolutions and spatial characteristics. 
This diversity enables a systematic assessment of how compression behavior varies across smooth, structured, and highly textured content. 
All images are processed in grayscale prior to compression and quantum state preparation.
For datasets originally provided in RGB format, we retain only the luminance component, ensuring consistency with the JPEG compression pipeline and avoiding color-channel–dependent effects in the subsequent quantum encoding.
\begin{figure}[h]
	\centering
	\includegraphics[width=0.9\linewidth]{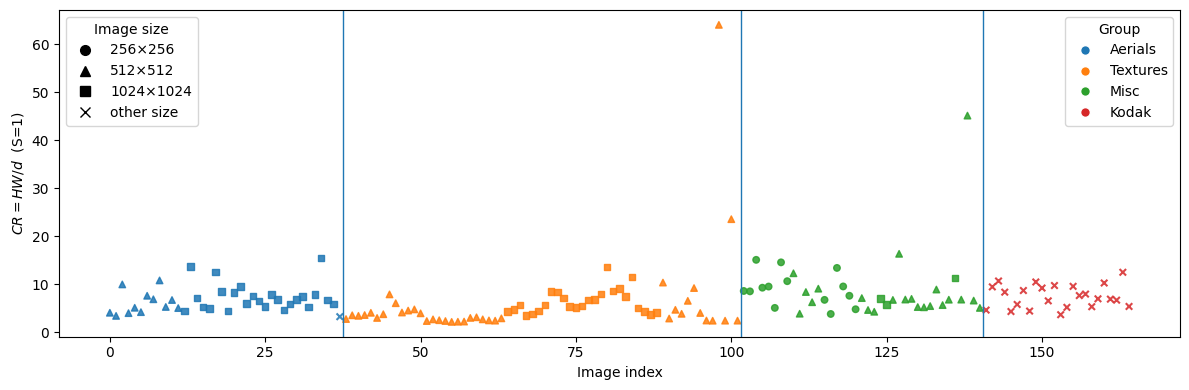}
	\caption{
		The CR obtained from JPEG compression ($S=1$) across standard image datasets.
		Each point corresponds to a single grayscale image from one of the Aerials, Textures, Miscellaneous, and Kodak sets.	
		Colors indicate image categories, marker shapes denote image resolutions, and crosses correspond to non-square images.
		While JPEG compression substantially reduces the number of active coefficients, most images remain exponentially dense in $n=\log_2 N$, with sparsity compatible with sparse quantum state preparation occurring only for highly structured or texture-dominated images.
	}
	\label{fig:Compression_ratio}
\end{figure}

For an image of size $N = HW$, we define the compression ratio (CR) as
\begin{equation}
	\mathrm{CR} = \frac{N}{d},
\end{equation}
where $N = 2^n$ denotes the total number of (padded) pixels and $d$ is the number of non-zero coefficients after JPEG compression.
Figure~\ref{fig:Compression_ratio} reports the observed CRs across the benchmark suite, grouped by image category and resolution.
Specifically, the measured CR ranges are
\begin{itemize}
	\item \textbf{Aerials}: 38 images, $\mathrm{CR}\in[3.23,\,15.43]$,
	\item \textbf{Textures}: 64 images, $\mathrm{CR}\in[2.15,\,64.00]$,
	\item \textbf{Miscellaneous}: 39 images, $\mathrm{CR}\in[3.79,\,45.10]$,
	\item \textbf{Kodak}: 24 images, $\mathrm{CR}\in[3.64,\,12.48]$.
\end{itemize}
Across all datasets, JPEG compression substantially reduces the number of active coefficients relative to the pixel domain.
However, for the majority of natural images, the resulting representations remain exponentially dense in $n=\log_2 N$.
Indeed, for an image with $N=2^n$ pixels, JPEG decomposes the image into $N/64 = 2^{n-6}$ blocks, and even in the most favorable case where only the DC coefficient of each block is retained, the total number of non-zero coefficients still scales as $2^{n-6}$.
Consequently, JPEG quantization alone is generally insufficient to guarantee sparsity compatible with SQSP, and in the worst case the resulting representations remain dense, necessitating the use of exact amplitude encoding.
\begin{figure}[h]
	\centering
	\includegraphics[width=0.9\linewidth]{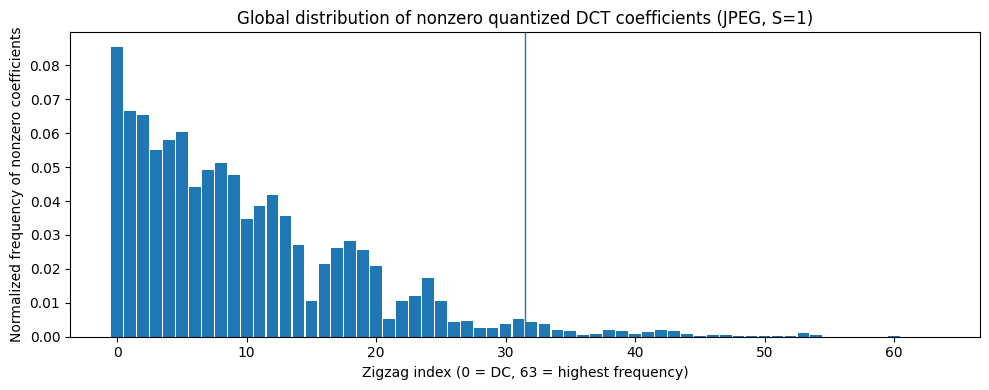}
	\caption{Global distribution of non-zero quantized DCT coefficients across zigzag positions.
		The histogram aggregates all images after partitioning into $8\times 8$ blocks and applying JPEG quantization with the standard luminance matrix ($S=1$).
		Each bar reports the normalized frequency with which a given zigzag index contains a non-zero coefficient.
		The strong concentration at low-frequency zigzag indices indicates that, while images may remain dense overall, the non-zero support within each block is predominantly confined to the first few zigzag positions.}
	\label{fig:histogram}
\end{figure}

Although JPEG does not typically produce a globally sparse representation, and the block partitioning distributes non-zero coefficients across the full image, the \emph{intra-block} support exhibits a highly structured pattern that can be exploited for QSP.
Figure~\ref{fig:histogram} shows the normalized distribution of non-zero quantized DCT coefficients over zigzag positions, aggregated across all images and all $8\times 8$ blocks.

The concentration of non-zero coefficients in low-frequency zigzag indices has a direct algorithmic consequence: it enables a \emph{controlled truncation} of the coefficient space within each $8\times 8$ block. 
A full block requires six qubits to address all 64 coefficients, whereas restricting to the first 32 zigzag coefficients reduces this to five qubits. 
Since the cost of exact amplitude encoding scales with the addressed Hilbert space dimension, truncation yields a proportional reduction in entangling gates while preserving perceptual reconstruction quality, as later quantified by PSNR and SSIM. 
Moreover, the effective number of significant coefficients is image-dependent, and in many blocks substantially fewer than 32 coefficients suffice. 
This motivates hybrid strategies in which the number of retained coefficients (and thus the quantum resources allocated per block) is determined by the observed coefficient distribution rather than fixed a priori.

\subsection{JQPIE Method}\label{sec:jqpie_method}

JQPIE is a hybrid preparation scheme that follows the JPEG logic within an amplitude-encoding (QPIE) setting. 
Instead of loading pixel intensities directly, JQPIE prepares a quantum state whose amplitudes encode the JPEG quantized coefficients, exploiting the fact that JPEG concentrates most of the signal energy into low-frequency components.
Concretely, the method begins from the zigzag-ordered quantized vector $\hat{Z}_j$ in Eq.~\ref{eq:Z_j_hat}, rather than the spatial-domain pixel block $B_j$.

Following the HQSP workflow (Sec.~\ref{sec:hybrid_QSP}), after classical compression of the image $I$, the JQPIE procedure proceeds in two quantum stages:
\begin{enumerate}
	\item QSP of the compressed coefficient representation, and
	\item coherent quantum decompression to reconstruct the image state.
\end{enumerate}
The remainder of this section details both stages and analyzes their associated quantum resource requirements.

\subsubsection{QSP of Compressed Image}\label{sec:QSP}
\begin{figure}[t]
	\centering
	\begin{subfigure}[t]{0.36\linewidth}
		\centering
		\includegraphics[width=\linewidth]{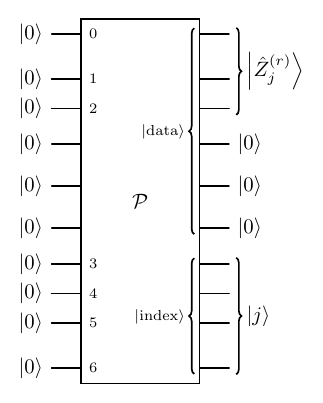}
		\caption{}
		\label{fig:preparation_jqpie}
	\end{subfigure}
	\begin{subfigure}[t]{0.4\linewidth}
		\centering
		\includegraphics[width=\linewidth]{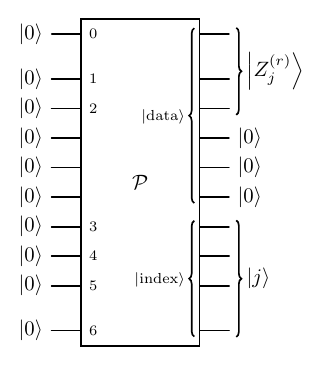}
		\caption{}
		\label{fig:preparation_qf-jqpie}
	\end{subfigure}
	\caption{
		Quantum state preparation (QSP) stage for the proposed hybrid image preparation schemes.
		In both circuits, the data and index registers are initialized in the $\ket{0}$ state.
		(a)~\textbf{JQPIE.} The state-preparation operator $\mathcal{P}$ loads the truncated, zigzag-ordered \emph{quantized} DCT coefficients $\hat{Z}_j^{(r)}$ on the quantum registers.		
		(b)~\textbf{QF-JQPIE.} The same preparation structure is used to encode the truncated zigzag-ordered \emph{unquantized} DCT coefficients $Z_j^{(r)}$.
	}
	\label{fig:preparation_circuit}
\end{figure}
To prepare JPEG-compressed image data as amplitudes of a quantum state, the original QPIE formulation must be adapted to the block-wise structure imposed by the $8\times 8$ DCT.
In particular, JPEG processing introduces a natural separation between \emph{block indices} and \emph{intra-block coefficients}, which must be reflected explicitly in the quantum register layout.

Let the input image be partitioned into $8\times8$ blocks. Equation~\ref{eq:QPIE-2D} can then be rewritten in block form as
\begin{equation}
	\ket{I}
	=
	\sum_{j=0}^{2^{h+w-6}-1}
	\ket{j}\ket{B_j},
	\qquad
	\ket{B_j}
	=
	\frac{1}{\sqrt{\sum_{x,y=0}^{7} X_j(x,y)^2}}
	\sum_{x=0}^{7}\sum_{y=0}^{7}
	X_j(x,y)\ket{x}\ket{y},
\end{equation}
where $\ket{j}$ denotes the \emph{index register} labeling the image blocks, and $\ket{B_j}$ is a six-qubit \emph{data register} encoding the normalized pixel intensities within block $j$.

Since the DCT coefficients within each $8\times8$ JPEG block are vectorized into a one-dimensional ordering, it is convenient to express each block state in linearized form,
\begin{equation}
	\ket{B_j}
	=
	\frac{1}{\sqrt{\sum_{k=0}^{63} X_j(k)^2}}
	\sum_{k=0}^{63} X_j(k)\ket{k},
	\qquad
	k = 8x + y.
\end{equation}
After quantization and zigzag reordering, the compressed block is represented by the coefficient vector $\hat{Z}_j$, which can be encoded as
\begin{equation}
	\ket{\phi}
	=
	\sum_{j=0}^{2^{h+w-6}-1}
	\ket{j}\ket{\hat{Z}_j},
	\qquad
	\ket{\hat{Z}_j}
	=
	\frac{1}{\sqrt{\sum_{k=0}^{63} \hat{C}_j(k)^2}}
	\sum_{k=0}^{63}
	\hat{C}_j\!\big(\pi(k)\big)\ket{k}.
\end{equation}

The vast majority of natural images remain dense after JPEG quantization.
Although quantization eliminates many raw pixel intensities, the block-wise partitioning distributes non-zero coefficients across the coefficient space, preventing direct sparsity-based speedups.
Consequently, applying exact QSP directly to $\ket{\hat{Z}_j}$ does not reduce the exponential preparation cost.

Nevertheless, the zigzag ordering introduces a highly structured distribution of coefficients: the DC component and low-frequency AC coefficients appear first, while high-frequency coefficients are typically negligible.
Motivated by the distribution shown in Fig.~\ref{fig:histogram}, we introduce a truncated representation that retains only the leading $2^{r}$ coefficients per block,
\begin{equation}
	\ket{\phi_1}
	=
	\sum_{j=0}^{2^{h+w-6}-1}
	\ket{j}
	\left(
	\ket{0}^{\otimes \ell}
	\otimes
	\frac{1}{A_j}
	\sum_{k=0}^{2^{r}-1}
	\hat{C}_j\!\big(\pi(k)\big)\ket{k}
	\right)
	=
	\sum_{j=0}^{2^{h+w-6}-1}
	\ket{j}
	\left(
	\ket{0}^{\otimes \ell}
	\otimes
	\ket{\hat{Z}_j^{(r)}}
	\right),
	\quad
	r \in \{2,3,4,5\},
\end{equation}
where $A_j=\sqrt{\sum_{k=0}^{2^r-1} \hat{C}_j(\pi(k))^2}$.
Here, $\ket{\hat{Z}_j^{(r)}}$ denotes the normalized truncated zigzag-ordered coefficient state containing the first $2^{r}$ coefficients of block $j$, and $\ell = 6 - r$ denotes the number of \emph{inactive} data-register qubits fixed in the $\ket{0}$ state during preparation.

The truncated state $\ket{\phi_1}$ remains dense and therefore must be prepared using an exact amplitude-encoding circuit~\cite{iten2016,mottonen2004quantum,sun2023}.
This preparation can be written as
\begin{equation}\label{eq:QSP_jqpie_P_operator}
	\ket{\phi_1}
	=
	\mathcal{P}
	\bigl(\ket{0}^{\otimes (h+w-6)}\bigr)_{\mathrm{index}}
	\bigl(\ket{0}^{\otimes \ell}\ket{0}^{\otimes r}\bigr)_{\mathrm{data}},
\end{equation}
where $\mathcal{P}$ acts on the $(h+w-\ell)$ active qubits spanning the index and data registers.
By excluding $\ell$ data qubits from the preparation stage, the dominant preparation cost is reduced to
$\mathcal{O}(2^{h+w-\ell})$, which constitutes the main gate complexity of the JQPIE algorithm.
Although this reduction is constant-factor relative to full QPIE, it yields a substantial practical decrease in circuit depth for near-term devices.

Figure~\ref{fig:preparation_jqpie} illustrates the resulting QSP circuit for a $2^{5}\times2^{5}$ image with truncation level $r=3$. During the preparation stage, only the active data qubits participate in amplitude encoding, while the inactive qubits remain fixed in the $\ket{0}$ state. The inactive qubits subsequently become entangled with the rest of the data register during the quantum inverse zigzag transformation, which restores the original frequency ordering.

\subsubsection{Quantum Inverse Zigzag Transform}\label{sec:quantum_zigzag}

The first step of the quantum decompression stage is to restore the original frequency-domain ordering within each $8\times8$ block by undoing the zigzag serialization applied during classical JPEG compression.
This task is accomplished by the zigzag permutation operator $P$, which acts on computational basis states as
\begin{equation}
	P\ket{k} = \ket{\pi(k)}.
\end{equation}
Because $P$ is a permutation matrix, it is unitary and admits an exact inverse given by $P^\dagger=P$.

Applying the inverse zigzag operator to the compressed quantum state $\ket{\phi_1}$ yields
\begin{equation}
	\ket{\phi_2} = \big(\mathbb{I}^{\otimes(h+w-6)} \otimes P^\dagger\big)\ket{\phi_1}
	= \sum_{j=0}^{2^{h+w-6}-1}\ket{j}
	\left(\frac{1}{A_j}
	\sum_{k=0}^{2^r-1} \hat{C}_j\!\big(\pi(k)\big)\ket{\pi(k)}
	\right)= \sum_{j=0}^{2^{h+w-6}-1}\ket{j}\ket{\psi_j}.
\end{equation}
Here, $\mathbb{I}$ denotes the single-qubit identity operator, and $P^\dagger$ restores the original two-dimensional frequency ordering of the quantized DCT coefficients defined in Eq.~\ref{eq:V_j_hat}, while preserving all amplitudes exactly.
This step prepares the coefficients in the appropriate layout for the subsequent inverse quantization and inverse DCT operations.

\paragraph{\textbf{Quantum Circuit Implementation}:}
From a resource perspective, implementing a dense arbitrary basis permutation on $n$ qubits generally requires an exponential number of elementary quantum gates~\cite{khandelwal2024classification,soeken2019compiling}.
In the present setting, however, the inverse zigzag operator $P$ acts only on the $6$-qubit data register associated with each $8\times8$ block and therefore constitutes a constant-size operation whose complexity does not scale with the overall image resolution.
Moreover, within the JQPIE framework only a truncated set of low-frequency zigzag coefficients is populated, while all remaining basis states are fixed in the $\ket{0}$ state.
This structure allows the full inverse zigzag permutation to be replaced by a truncated permutation that acts non-trivially only on the retained coefficients, leading to a further reduction in gate count and circuit depth.
The explicit construction and circuit-level realization of this truncated permutation operator are presented in Appendix~\ref{sec:appendix}.

\subsubsection{Quantum Inverse Quantization}\label{sec:quantum_quantization}
\begin{figure}[t!]
	\centering	
	\begin{subfigure}{0.75\linewidth}
		\centering
		\includegraphics[width=\linewidth]{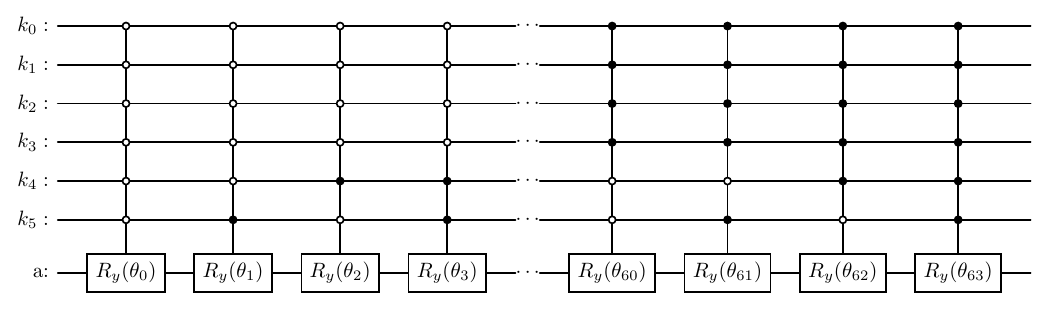}
		\caption{}
		\label{fig:quantization_circuit_a}
	\end{subfigure}
	
	\vspace{1em}
	
	\begin{subfigure}{0.95\linewidth}
		\centering
		\includegraphics[width=\linewidth]{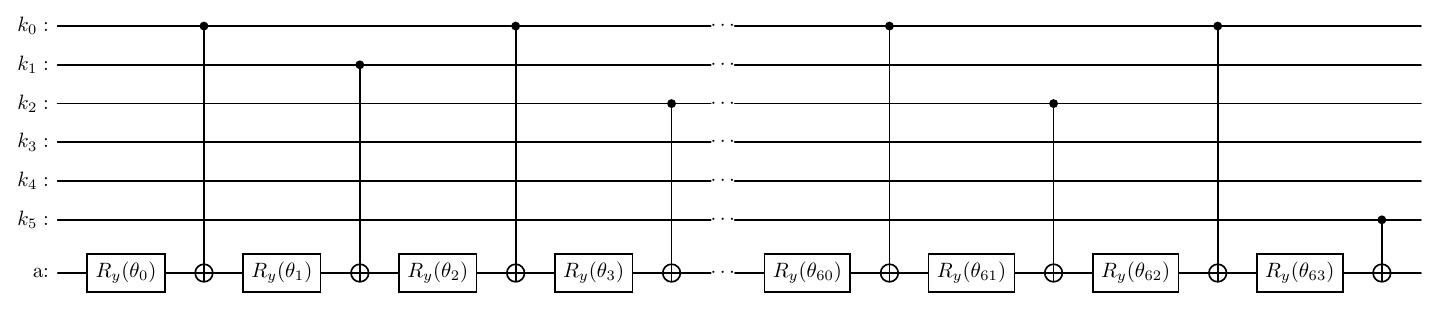}
		\caption{}
		\label{fig:quantization_circuit_b}
	\end{subfigure}
	\caption{Quantum implementation of the block-encoded quantization matrix.  
		(a)~The ancilla qubit receives uniformly controlled rotations conditioned on the 6-bit index $k$.  
		(b)~Equivalent decomposition using the Möttönen construction~\cite{mottonen2004transformation}.}
	\label{fig:quantization_circuit_both}
\end{figure}
The inverse quantization step rescales each quantized DCT coefficient by the corresponding entry of the quantization matrix
$Q \in \mathbb{R}^{8\times8}$ (see Eq.~\ref{eq:quantization_matrix}).
At the block level, this operation is expressed as an element-wise multiplication,
\begin{equation}
	C_j = Q \odot \hat{C}_j ,
\end{equation}
where $\odot$ denotes the element-wise product.
Equivalently, after vectorization, inverse quantization can be written as the action of a diagonal operator,
\begin{equation}
	V_j = \tilde{Q}\,\hat{V}_j ,
	\qquad
	\tilde{Q} = \mathrm{diag}\!\big(Q(0), \ldots, Q(k), \ldots, Q(63)\big),
	\qquad k = 8u + v .
\end{equation}

Since the diagonal operator \(\tilde{Q}\) is generally non-unitary ($\tilde{Q}^\dagger \neq \tilde{Q}^{-1}$), it cannot be implemented directly as a quantum operation. To realize inverse quantization coherently, we therefore employ a block-encoding construction in which a rescaled version of \(\tilde{Q}\) appears as a sub-block of a larger unitary operator acting on an ancillary qubit. 
To this end, we first normalize the quantization coefficients such that all entries lie in the interval \([0,1]\),
\begin{equation}
	\lambda = \max_k Q(k), \qquad d_k = \frac{Q(k)}{\lambda},
\end{equation}
and define the rescaled diagonal operator \(\tilde{D} = \mathrm{diag}(d_0,\ldots,d_{63})\), which satisfies \(\|\tilde{D}\| \leq 1\). In this form, the action of inverse quantization on the computational basis states is given explicitly by
\begin{equation}
	\tilde{D}\ket{k} = d_k \ket{k} = \frac{Q(k)}{\lambda}\ket{k}.
\end{equation}
Each coefficient \(d_k\) is implemented using a controlled rotation on an ancillary qubit with rotation angle \(\theta_k = 2\arccos(d_k)\), resulting in the single-qubit operation
\begin{equation}
	R_y(\theta_k) =
	\begin{pmatrix}
		d_k & -\sqrt{1-d_k^2} \\
		\sqrt{1-d_k^2} & d_k
	\end{pmatrix}.
\end{equation}
Collectively, these controlled rotations define a block-encoded unitary operator of the form
\begin{equation}
	U_Q =
	\begin{pmatrix}
		\tilde{D} & -\sqrt{\mathbb{I}_{D}-\tilde{D}^2} \\
		\sqrt{\mathbb{I}_{D}-\tilde{D}^2} & \tilde{D}
	\end{pmatrix}
	=
	\mathbb{I}\otimes \tilde{D}
	-
	i\sigma_y \otimes \sqrt{\mathbb{I}_{D}-\tilde{D}^2},
\end{equation}
where $\mathbb{I}_{D}$ denotes the identity operator acting on the same Hilbert space as $\tilde{D}$.
By construction, $\|\tilde{D}\|\le 1$, ensuring that $\mathbb{I}_{D}-\tilde{D}^2$ is positive semidefinite.
The upper-left block of $U_Q$ implements the rescaled inverse-quantization operator $\tilde{D} = \tilde{Q}/\lambda$,
while the remaining blocks ensure the overall unitarity of the transformation $(U_Q^\dagger = U_Q^{-1})$.

Since the block-encoded unitary $U_Q$ acts jointly on the data register and an ancillary qubit, we extend the state $\ket{\phi_2}$ by appending an ancilla initialized in the $\ket{0}_a$ state and apply $U_Q$ to the joint system, yielding
\begin{eqnarray}
	\ket{\phi_3} &=& \left(\mathbb{I}\otimes \mathbb{I}^{\otimes(h+w-6)} \otimes \tilde{D} -i\sigma_y \otimes \mathbb{I}^{\otimes(h+w-6)} \otimes \sqrt{\mathbb{I}_{\texttt{D}}-\tilde{D}^2}\right) \ket{0}_a\ket{\phi_2}	\nonumber\\
	&=&
	\ket{0}_a \sum_j \ket{j}\Tilde{D}\ket{\psi_j} + \ket{1}_a \sum_j \ket{j}\sqrt{\mathbb{I}_{\texttt{D}} - \Tilde{D}^2}\ket{\psi_j}.
\end{eqnarray}
The desired output resides in the $\ket{0}_a$ subspace, expanding the $\ket{0}_a$ contribution gives
\begin{eqnarray}
	\Tilde{D} \ket{\psi_j}	=
	\frac{1}{A_j}
	 \sum_{k=0}^{2^r-1} \hat{C}_j(\pi(k))\Big(\Tilde{D}\ket{\pi(k)}\Big) = \frac{1}{A_j\lambda}
	\sum_{k=0}^{2^r-1} \hat{C}_j(\pi(k)) Q(\pi(k))\ket{\pi(k)}= \frac{1}{A_j\lambda}
	\sum_{k=0}^{2^r-1} {C}_j(\pi(k)) \ket{\pi(k)}.
\end{eqnarray}
Finally, the one-dimensional index $k$ is mapped back to its two-dimensional frequency coordinates,
\begin{equation}
	\sum_{k=0}^{63} C_j(\pi(k))\ket{\pi(k)} 
	= \sum_{u=0}^{7}\sum_{v=0}^{7} C_j(u,v)\ket{u}\ket{v}, \qquad u = \left\lfloor \frac{\pi(k)}{8} \right\rfloor, 
	\qquad
	v = \pi(k) \bmod 8,
\end{equation}
which is proper for the two-dimensional inverse QDCT described in the next section.

\paragraph{\textbf{Quantum Circuit Implementation}:}
From an implementation perspective, the block-encoded inverse quantization is realized using a 6-fold uniformly controlled $R_y(\theta_k)$ rotation acting on a single ancilla qubit, controlled by the qubits in data register, as illustrated in Fig.~\ref{fig:quantization_circuit_a}.
Following the decomposition of Möttönen \emph{et al.}~\cite{mottonen2004transformation}, such a uniformly controlled rotation decomposes into $2^6$ \texttt{CX} gates and $2^6$ single-qubit rotations, with an overall circuit depth of $2^7+1$ (see Fig.~\ref{fig:quantization_circuit_b}).

\subsubsection{Quantum Inverse DCT}
The final step of the decompression phase applies the two-dimensional inverse quantum DCT to recover spatial-domain pixel amplitudes. Since the DCT is separable, the 2D transform is realized by applying the optimized 8-point inverse QDCT along rows and columns. The resulting inverse 2D-QDCT operator is
\begin{eqnarray}
	U^\dagger_{\mathrm{2D-QDCT}} = U^\dagger_{\mathrm{QDCT}} \otimes U^\dagger_{\mathrm{QDCT}}.
\end{eqnarray}
Applying this operation to the state $\ket{\phi_3}$ yields
\begin{eqnarray}
	\ket{\widetilde{I}} &=& \left(\mathbb{I}^{\otimes (h+w-5)} \otimes U^\dagger_{\mathrm{2D-DCT}} \right)\ket{\phi_3}, \nonumber\\
	&=& 
	\ket{0}_{a}  
	\Bigg[\frac{1}{\lambda}\sum_{j=0}^{2^{h+w-6}-1} \ket{j} \left( \frac{1}{A_j}
	\sum_{u=0}^{7}\sum_{v=0}^{7} \hat{C}_j(u,v) \left(U^\dagger_{\mathrm{QDCT}}\ket{u}\right) \left(U^\dagger_{\mathrm{QDCT}}\ket{v}\right)\right)\Bigg] + \ket{1}_{a}\ket{\text{junk}}, \nonumber\\
	&=& \ket{0}_{a}  \Bigg[\frac{1}{\lambda}\sum_{j=0}^{2^{h+w-6}-1} \ket{j} \left(\frac{1}{A_j}\sum_{x=0}^{7}\sum_{y=0}^{7} X_j(x,y)\ket{x}\ket{y}\right)\Bigg] + \ket{1}_{a}\ket{\text{junk}},
\end{eqnarray}
which encodes the reconstructed blocks in the desired spatial domain.
The useful image information resides in the $\ket{0}_a$ subspace; it can be isolated either by post-selection or, more efficiently by applying oblivious amplitude amplification (OAA)~\cite{berry2014exponential, kothari2014efficient} technique.

\paragraph{\textbf{Quantum Circuit Implementation}:}
To implement the inverse DCT coherently, we adopt the unitary circuit construction proposed by Tseng and Hwang~\cite{tseng2004quantum}, which realizes the fast DCT factorization~\cite{vetterli2003discrete} in a quantum-compatible form.
In this construction, an 8-point inverse QDCT acting on a three-qubit register requires $18$ \texttt{CX} gates and $33$ single-qubit rotations, with circuit depth $35$.
Since the two-dimensional transform is separable, the row and column inverse QDCTs act on independent registers.
As a result, the total gate count doubles, while the circuit depth remains unchanged.

\begin{figure}[t!]
	\centering
	\begin{subfigure}[t]{0.5\linewidth}
		\centering
		\includegraphics[width=\linewidth]{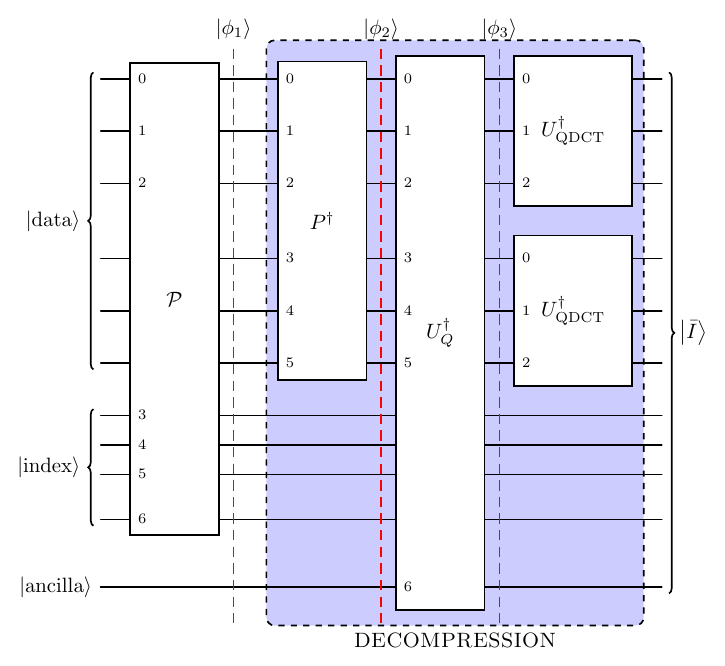}
		\caption{}
		\label{fig:full_quantum_circuit}
	\end{subfigure}
	\hfill
	\begin{subfigure}[t]{0.46\linewidth}
		\centering
		\includegraphics[width=\linewidth]{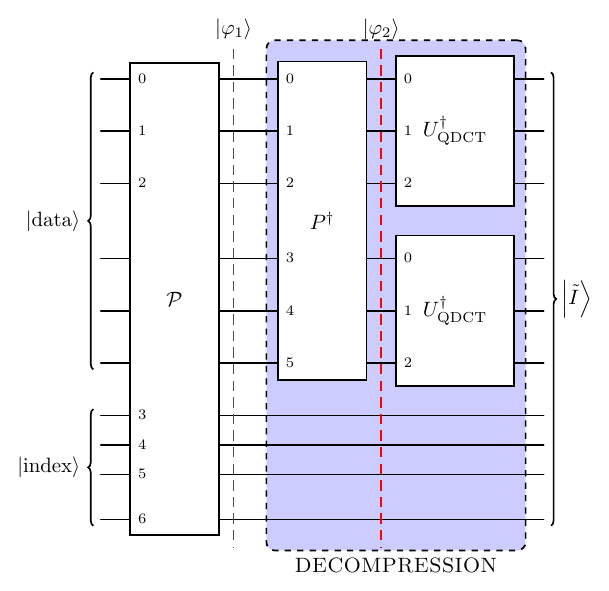}
		\caption{}
		\label{fig:full_quantum_circuit_qf}
	\end{subfigure}
	\caption{
		Illustrative quantum circuits for hybrid image preparation using JQPIE and QF-JQPIE for a sample $32\times32$ image.
		(a)~\textbf{JQPIE circuit.} The circuit consists of three registers: a data register, an index register, and a single ancillary qubit. 
		Image reconstruction is performed through four sequential stages.
		The operators $P^\dagger$ (inverse zigzag) and $U^\dagger_{\mathrm{QDCT}}$ (inverse quantum DCT) act exclusively on the data register; the numbers shown inside each block indicate the qubits involved in the corresponding operations.
		The state-preparation operator $\mathcal{P}$ acts jointly on the data and index registers, while the block-encoded inverse-quantization operator $U_Q^\dagger$ couples the data register to the ancillary qubit.
		Red dashed separators indicate the intermediate states $\ket{\phi_k}$ generated at each stage, as described in Sec.~\ref{sec:jqpie_method}.
		The final output state is $\ket{\Bar{I}}=\ket{0}_a\ket{\widetilde{I}}+\ket{1}_a\ket{\mathrm{junk}}$.
		(b)~\textbf{QF-JQPIE circuit.} The quantization-free variant uses only two registers (data and index) and consists of three stages.
		The circuit structure mirrors that of JQPIE but omits the inverse-quantization block and ancillary qubit.
		The intermediate states $\ket{\varphi_k}$ are defined in Sec.~\ref{sec:qf_jqpie}, and the final quantum output directly corresponds to the approximate image state $\ket{\widetilde{I}}$.
	}
	\label{fig:full_quantum_circuit_comparison}
\end{figure}

At this point, the coherent quantum realization of the JPEG decompression process is complete. Algorithm~\ref{alg:JQPIE} summarizes the full JQPIE approach. A schematic realization of the corresponding quantum circuit is shown in Fig.~\ref{fig:full_quantum_circuit}, where the individual stages of the method are explicitly illustrated.

\begin{algorithm}[h!]
	\caption{JQPIE}
	\label{alg:JQPIE}
	\begin{algorithmic}[1]
		
		\Require Image $I$, truncation level $r\in\{2,3,4,5\}$
		\Ensure Quantum state $\ket{\widetilde{I}}$ approximating the image $I$
		
		\Statex \textbf{Classical stage}
		\State Partition $I$ into $8\times 8$ blocks $B_j$.
		\State Compute 2D-DCT and quantize coefficients $\hat{C}_j(u,v)$.
		\State Apply zigzag permutation $P$ and keep first $2^r$ coefficients,
		\[
		\hat{Z}^{(r)}_j = \big[\hat{C}_j(\pi(0)),\, \hat{C}_j(\pi(1)),\, \ldots,\, \hat{C}_j(\pi(2^r-1))\big]^T,
		\]		
		
		\Statex \textbf{QSP}
		\State Prepare the compressed image $\hat{Z}^{(r)}_j$
		\[
		\ket{\phi_1} = \ket{0}_a\sum_{j=0}^{2^{h+w-6}-1} \ket{j} \left( \ket{0}^{\otimes\ell} 
		\frac{1}{A_j}\sum_{k=0}^{2^{r}-1} \hat{C}_j(\pi(k))\,\ket{k}\right)
		\]
		
		\Statex \textbf{Quantum Decompression}
		\State Inverse zigzag: \quad 
		\(
		\ket{\phi_2} = \big(\mathbb{I}^{\otimes(h+w-6)} \otimes P^\dagger\big)\ket{\phi_1}.
		\)
		\State Inverse quantization via block-encoding 
		\[
		\ket{\phi_3} = \left(\mathbb{I}\otimes \mathbb{I}^{\otimes(h+w-6)} \otimes \tilde{D} -i\sigma_y \otimes \mathbb{I}^{\otimes(h+w-6)} \otimes \sqrt{\mathbb{I}_{\texttt{D}}-\tilde{D}^2}\right)	\ket{0}_a \ket{\phi_2} 
		\]
		\State Apply 2D inverse QDCT: \quad
		\[
		\ket{\widetilde{I}} = \left(\mathbb{I}^{\otimes (h+w-5)} \otimes U^\dagger_{\mathrm{2D-DCT}}\right) \ket{\phi_3}.
		\]
		
		\State Post-select (or amplify) the $\ket{0}_a$ subspace.
		
	\end{algorithmic}
\end{algorithm}

\subsection{Quantization-Free JQPIE Method}\label{sec:qf_jqpie}

In classical JPEG compression, quantization of DCT coefficients reduces their numerical magnitude, which directly improves entropy coding efficiency by enabling shorter variable-length codes.
In contrast, quantum amplitude encoding does not rely on binary representations of coefficient values.
Instead, classical data are embedded as continuous amplitudes of a quantum state, where the numerical magnitude of a coefficient affects only the rotation angles used during state preparation, rather than the number of qubits, registers, or the asymptotic gate complexity of the circuit.

Motivated by the insensitivity of quantum amplitude encoding to coefficient magnitude, we introduce a quantization-free variant of JQPIE.
In this approach, high-frequency DCT components are removed by truncation in the zigzag-ordered domain, while the remaining coefficients are preserved in their original form.
This avoids the use of quantization matrices altogether and yields an ancilla-free state preparation procedure whose efficiency derives solely from frequency-domain sparsification.

As shown in Sec.~\ref{sec:hybrid_QIP}, the energy of most natural images is predominantly concentrated in the low-frequency region.
This allows the zigzag-ordered representation to be truncated without significant loss of visual fidelity.
The resulting truncated vectors can then be loaded into quantum registers using exact QSP methods with substantially reduced quantum resources.

Following the HQSP workflow, we first compressed the input image $I$ through the block-wise 2D DCT of non-overlapping $8\times 8$ blocks $B_j$ to obtain the corresponding frequency-domain blocks, $C_j$.
At this stage, unlike the JQPIE, instead of applying quantization, we directly perform a zigzag permutation that orders the coefficients from low to high spatial frequencies, yielding
\begin{equation}
	Z_j = P V_j = \big[C_j(\pi(0)), C_j(\pi(1)), \ldots, C_j(\pi(k)), \ldots, C_j(\pi(63))\big],
\end{equation}
where $V_j$ was defined in Eq.~\ref{eq:V_j}. 
Here, $Z_j(k) = C_j\big(\pi(k)\big)$ represents the $k$-th zigzag-ordered DCT coefficient of block $j$.

In the QF-JQPIE method, instead of loading the full zigzag-ordered coefficient vector $Z_j$, only the leading $2^{r}$ low-frequency coefficients of each block are prepared quantum mechanically.
Following the same state-preparation strategy as in the JQPIE formulation (cf.~Sec.~\ref{sec:QSP}), and using the operator $\mathcal{P}$ defined in Eq.~\ref{eq:QSP_jqpie_P_operator}, the QF-JQPIE variant prepares the state $\ket{\varphi_1}$ by loading the truncated zigzag-ordered coefficients $Z_j^{(r)}$ in place of the quantized coefficients $\hat{Z}_j^{(r)}$. 
Explicitly, the resulting state can be written as
\begin{equation}
	\ket{\varphi_1}
	=
	\sum_{j=0}^{2^{h+w-6}-1}
	\ket{j}
	\left(
	\ket{0}^{\otimes \ell}
	\frac{1}{K_j}
	\sum_{k=0}^{2^{r}-1}
	Z_j(k)\ket{k}
	\right)
	=
	\sum_{j=0}^{2^{h+w-6}-1}
	\ket{j}
	\left(
	\ket{0}^{\otimes \ell}
	\frac{1}{K_j}
	\sum_{k=0}^{2^{r}-1}
	C_j\!\big(\pi(k)\big)\ket{k}
	\right),
\end{equation}
where $K_j=\sqrt{\sum_{k=0}^{2^{r}-1} Z_j(k)^2}$ ensures proper normalization.
Here, the index register consists of $(h+w-6)$ qubits labeling the image blocks, while the data register contains six qubits corresponding to the $8\times8$ block structure.
This preparation step is schematically illustrated in Fig.~\ref{fig:preparation_qf-jqpie}.

To restore the original two-dimensional frequency ordering within each block, we apply the inverse zigzag permutation,
\begin{equation}
	\ket{\varphi_2}
	=
	(\mathbb{I}^{\otimes (h+w-6)} \otimes P^\dagger)\ket{\varphi_1}
	=	
	\sum_{j=0}^{2^{h+w-6}-1}
	\ket{j}
	\left(\frac{1}{K_j}
	\sum_{k=0}^{2^r-1}
	C_j\big(\pi(k)\big)\ket{\pi(k)}\right).
\end{equation}
Equivalently, this state may be expressed in two-dimensional frequency coordinates as
\begin{equation}
	\ket{\varphi_2}
	=	
	\sum_{j=0}^{2^{h+w-6}-1}
	\ket{j}
	\left(\frac{1}{K_j}
	\sum_{u=0}^{7}\sum_{v=0}^{7}
	C_j(u,v) \ket{u}\ket{v}\right)	,\qquad u = \left\lfloor \frac{\pi(k)}{8} \right\rfloor, 
	\qquad
	v = \pi(k) \bmod 8.
\end{equation}
Finally, the spatial-domain pixel amplitudes are reconstructed by applying the inverse two-dimensional quantum DCT,
\begin{equation}
	\ket{\widetilde{I}}
	=
	\left(
	\mathbb{I}^{\otimes (h+w-6)} \otimes U_{\mathrm{2D\text{-}DCT}}^\dagger
	\right)
	\ket{\varphi_2}
	=	
	\sum_{j=0}^{2^{h+w-6}-1}	
	\ket{j}
	\left(\frac{1}{K_j}
	\sum_{x=0}^{7}\sum_{y=0}^{7}
	X_j(x,y)\ket{x}\ket{y}\right).
\end{equation}

The resulting state $\ket{\widetilde{I}}$ encodes an approximation of image in the QPIE representation. 
A full quantum circuit implementation of the QF-JQPIE method is illustrated in Fig.~\ref{fig:full_quantum_circuit_qf}, and the associated algorithmic procedure is provided in Algorithm~\ref{alg:QF_JQPIE}.

\begin{algorithm}[h!]
	\caption{QF-JQPIE: Quantization-Free JPEG-Assisted Quantum Image Preparation}
	\label{alg:QF_JQPIE}
	\begin{algorithmic}[1]
		
		\Require Image $I$, truncation level $r \in \{2,3,4,5\}$
		\Ensure Quantum state $\ket{\widetilde{I}}$ approximating the image $I$
		
		\Statex \textbf{Classical pre-processing}
		\State Partition $I$ into $8\times8$ blocks $B_j$.
		\State Compute the 2D DCT of each block to obtain coefficients $C_j(u,v)$.
		\State Apply zigzag permutation $P$ and retain the first $2^r$ coefficients:
		\[
		Z_j^{(r)} =
		\big[C_j(\pi(0)),\, C_j(\pi(1)),\, \ldots,\, C_j(\pi(2^r-1))\big]^T.
		\]
		
		\Statex \textbf{QSP}
		\State Prepare the state $Z_j^{(r)}$
		\[
		\ket{\varphi_1}
		=
		\sum_{j=0}^{2^{h+w-6}-1}
		\ket{j}		
		\left(
		\ket{0}^{\otimes \ell}		
		\frac{1}{K_j}\sum_{k=0}^{2^r-1} Z_j^{(r)}(k)\ket{k}
		\right).
		\]
		
		\Statex \textbf{Quantum Decompression}
		\State Apply inverse zigzag reordering:
		\[
		\ket{\varphi_2}
		=
		(\mathbb{I}^{\otimes (h+w-6)} \otimes P^\dagger)\ket{\varphi_1}.
		\]
		\State Apply the inverse 2D quantum DCT:
		\[
		\ket{\widetilde{I}}
		=
		(\mathbb{I}^{\otimes (h+w-6)} \otimes U_{\mathrm{2D\text{-}DCT}}^\dagger)
		\ket{\varphi_2}.
		\]		
	\end{algorithmic}
\end{algorithm}


\section{Simulation Results and Discussion}\label{sec:results_discussion}

\begin{figure}[h]
	\centering
	\includegraphics[width=0.9\linewidth]{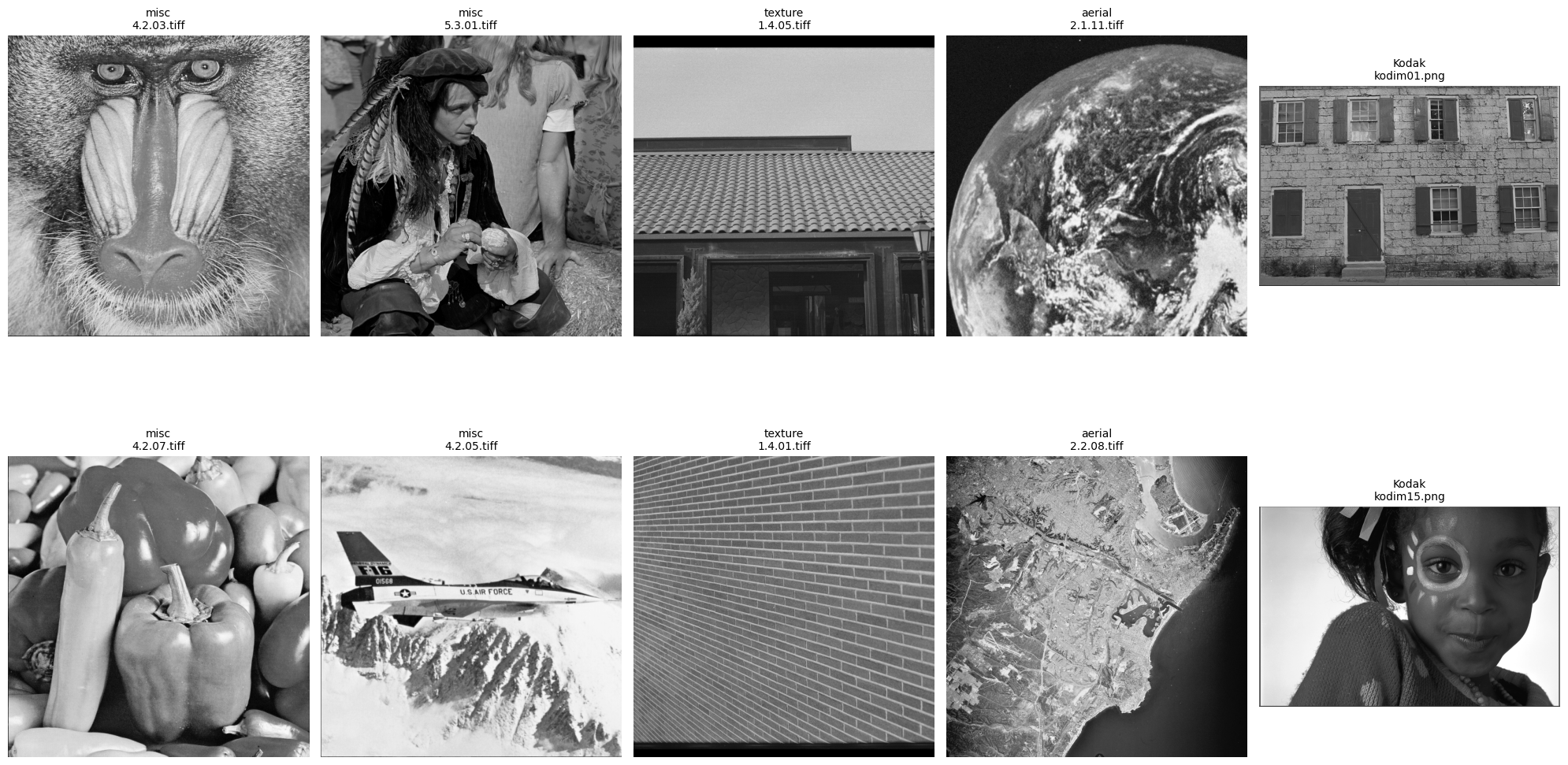}
	\caption{
		Representative images selected from the USC--SIPI and Kodak datasets, covering diverse spatial structures and texture characteristics across different image classes.
	}
	\label{fig:benchmark_images}
\end{figure}

This section reports quantum simulation results for the two proposed hybrid quantum image preparation schemes.
All simulations were performed using the \texttt{Qiskit} framework~\cite{javadi2024quantum}, following the circuit constructions described in Sec.~\ref{sec:hybrid_QIP}.
We employ statevector simulation, which provides direct access to the ideal quantum state prepared by the circuit prior to measurement.
This setting enables an exact assessment of the intrinsic quality of the prepared quantum image states, free from hardware noise and finite-shot sampling effects.

In the amplitude-encoding paradigm, a quantum image is represented by a normalized state $\ket{\Psi}=\sum_k c_k\ket{k}$, where the amplitudes encode normalized pixel intensities.
In a physical implementation, repeated projective measurements yield outcome probabilities proportional to $|c_k|^2$, from which pixel values are reconstructed by taking square roots and applying classical rescaling.
This rescaling includes the normalization factor introduced during state preparation and, in the JQPIE framework, an additional multiplicative factor $\lambda$ arising from the inverse-quantization step.

We evaluate Algorithms~\ref{alg:JQPIE} and~\ref{alg:QF_JQPIE} using benchmark images introduced in Sec.~\ref{sec:hybrid_QIP}, covering a broad range of spatial structures.
Reconstruction quality is quantified using the predefined PSNR and SSIM metrics (see Sec.~\ref{sec:metrics}) after the quantum inverse transformations.
The reduction in quantum resources, measured in terms of \texttt{CX} gate count and circuit depth, is fully determined by the truncation parameter $r$.
Specifically, retaining $2^r$ coefficients per $8\times8$ block reduces the QSP cost by a factor scaling proportionally with $2^{-\ell}$ relative to the full QPIE construction.
As a result, the reduction in \texttt{CX} gate count and circuit depth is directly controlled by the truncation level $r$ and is independent of image size.
For this reason, explicit gate counts are not reported here, as the resource reduction follows deterministically from the chosen truncation parameter.
\begin{figure}[h]
	\centering
	\includegraphics[width=0.9\linewidth]{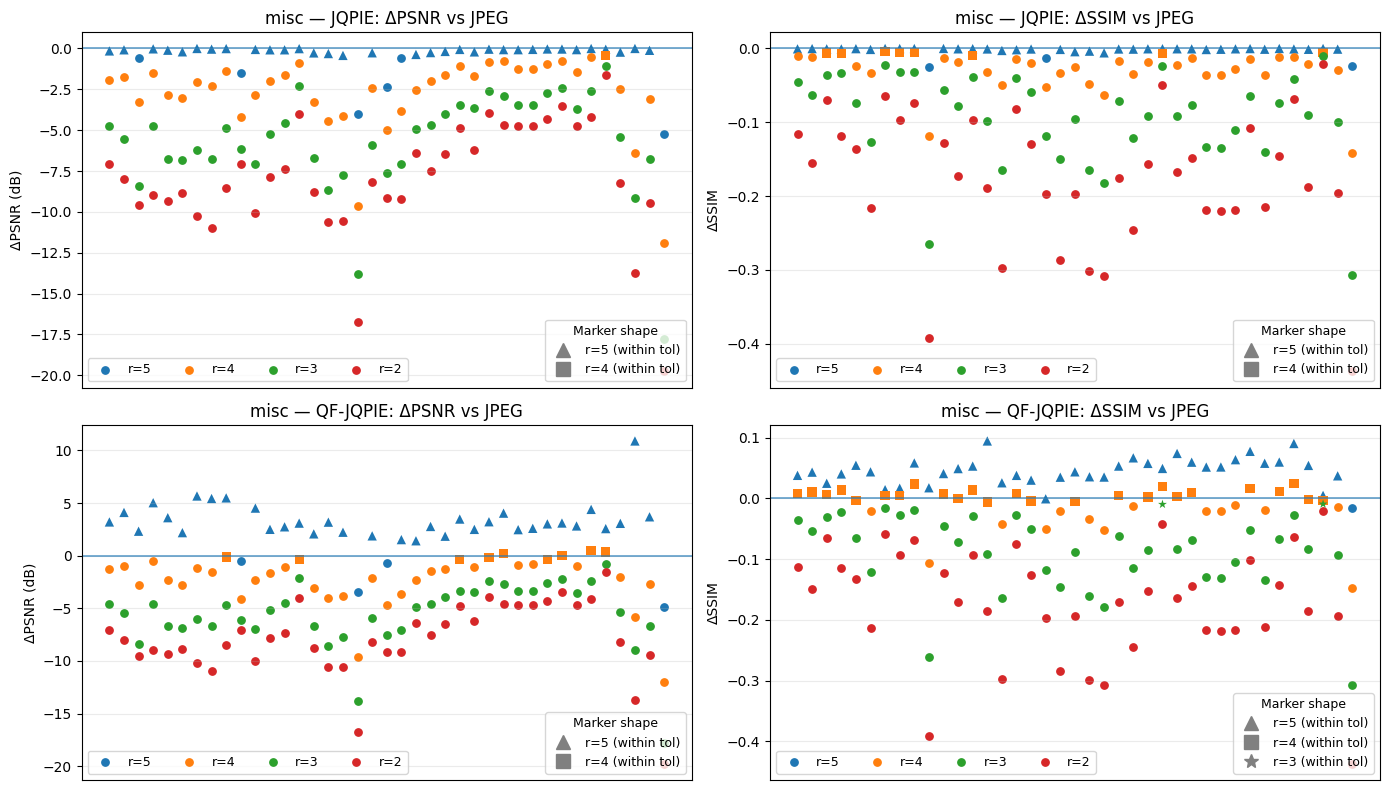}
	\caption{
		Reconstruction quality on the miscellaneous image dataset for the proposed hybrid image preparation schemes.
		Each subplot reports deviations relative to the classical JPEG decoding ($S=1$), with $\Delta$PSNR and $\Delta$SSIM measured after quantum reconstruction.
		Top row: JQPIE, where coefficient quantization and truncation lead to a systematic degradation relative to the JPEG baseline.
		Bottom row: QF-JQPIE, where truncation in the orthonormal DCT domain frequently preserves or improves reconstruction quality.
		Each point corresponds to a single grayscale image, evaluated at truncation levels $r\in\{5,4,3,2\}$, retaining the first $2^r$ low-frequency zigzag-ordered coefficients per $8\times8$ block. The horizontal reference line ($\Delta=0$) indicates the JPEG baseline.
	}
	\label{fig:PSNR_SSIM_JQPIE_&_QF_misc}
\end{figure}

We first analyze the miscellaneous image category, which exhibits substantial diversity in spatial structure and content.
Representative examples drawn from the USC--SIPI and Kodak datasets, including miscellaneous images together with their original labels, are shown in Fig.~\ref{fig:benchmark_images}.
For each image, reconstruction quality is evaluated at multiple truncation levels $r$, corresponding to retaining the first $2^{r}$ zigzag-ordered DCT coefficients per $8\times8$ block.
Figure~\ref{fig:PSNR_SSIM_JQPIE_&_QF_misc} summarizes the reconstruction quality relative to classical JPEG decoding for both hybrid schemes.
In these plots, the horizontal reference level corresponds to the classical JPEG baseline ($S=1$), and points closer to this level indicate reconstructions that are visually and quantitatively comparable to classical JPEG.

For JQPIE (top row), the reconstruction quality of the quantum-prepared images degrades systematically as truncation becomes more aggressive.
At $r=5$, the majority of images remain close to the JPEG baseline in both PSNR and SSIM, indicating limited perceptual degradation while achieving a $50\%$ reduction in \texttt{CX} gate count and circuit depth during state preparation.
As truncation increases ($r<5$), reconstruction fidelity deteriorates progressively.
Due to the combined effects of coefficient quantization and truncation, all JQPIE reconstructions eventually fall below the JPEG baseline for sufficiently small $r$.

\begin{figure}[h]
	\centering
	\includegraphics[width=0.9\linewidth]{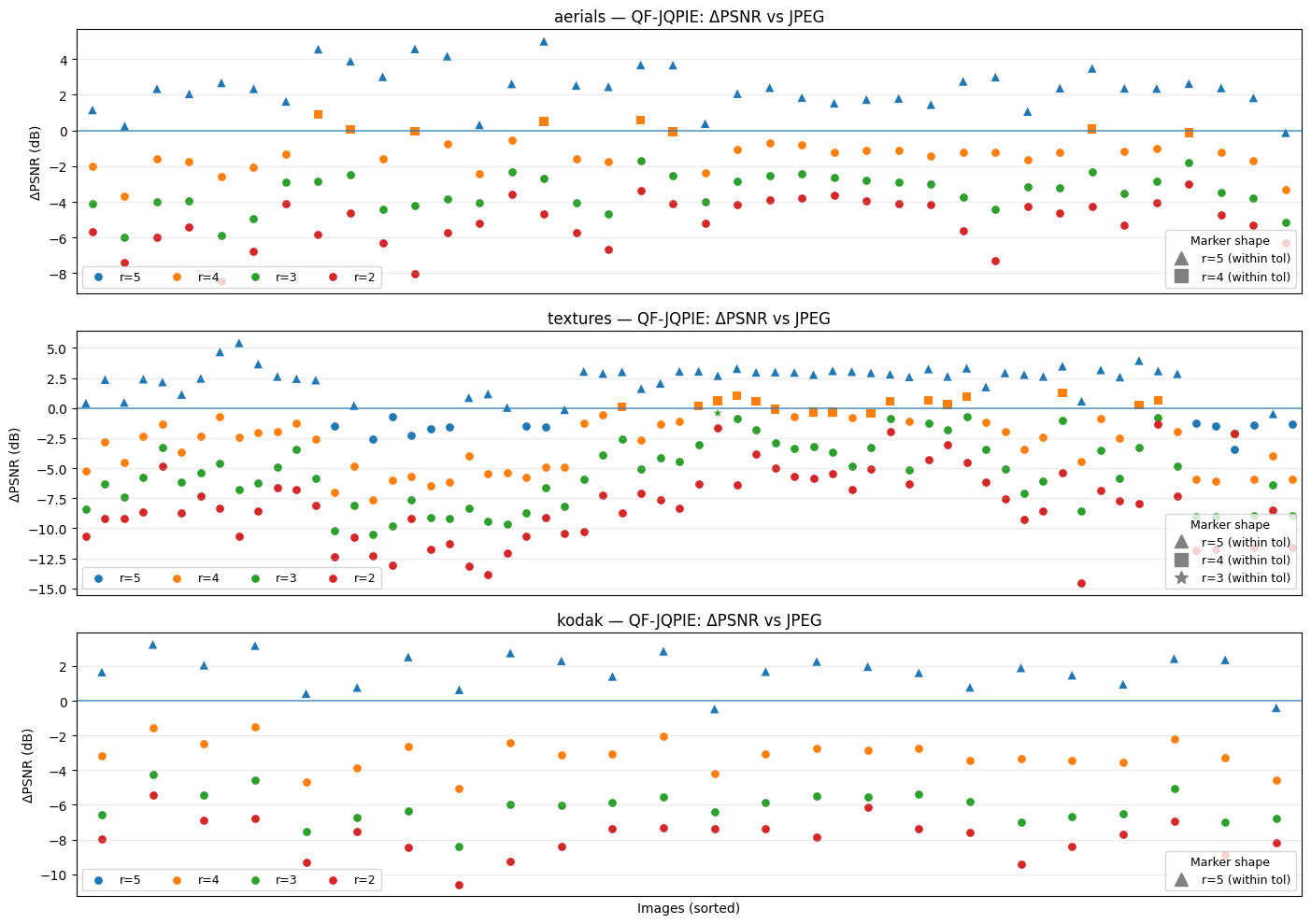}
	\caption{
		Reconstruction quality of the quantization-free JPEG-assisted QPIE (QF-JQPIE) scheme measured by
		$\Delta$PSNR relative to the classical JPEG decoding ($S=1$) for the aerial, texture, and Kodak image
		datasets.
		Each point corresponds to a single grayscale image, and the horizontal axis indexes images in
		sorted order within each dataset.
		Positive values indicate reconstruction quality exceeding the JPEG baseline.
		The uniform horizontal distribution emphasizes dataset-level trends rather than image ordering,
		highlighting the robustness of QF-JQPIE across diverse image classes.
	}
	\label{fig:PSNR_JQPIE_&_QF_others}
\end{figure}

For the miscellaneous dataset (39 images in total), we quantify reconstruction performance using tolerance thresholds of
$\Delta\mathrm{PSNR} \geq -0.5\,\mathrm{dB}$ and $\Delta\mathrm{SSIM} \geq -0.01$.
At $r=5$, JQPIE reconstructs approximately $85\%$ of the images with PSNR values indistinguishable from classical JPEG, while the remaining images exhibit at most a $2\,\mathrm{dB}$ PSNR reduction.
Using the SSIM criterion, $92\%$ of the images preserve structural similarity comparable to the JPEG baseline.

It is important to note that PSNR and SSIM capture complementary aspects of reconstruction quality.
While PSNR reflects pixel-wise fidelity and is particularly sensitive to quantization and high-frequency distortions, SSIM emphasizes the preservation of structural and perceptual content.
As a result, reconstructions satisfying at least one tolerance criterion may still appear visually comparable to JPEG, whereas reconstructions failing both metrics typically exhibit noticeable degradation.

As truncation becomes more aggressive, the fraction of images satisfying the PSNR tolerance decreases rapidly.
As shown in Fig.~\ref{fig:PSNR_JQPIE_&_QF_others} (top row), at $r=4$ only a single image remains within the PSNR tolerance, while $8$ images continue to satisfy the SSIM criterion.
For these cases, JQPIE achieves a $75\%$ reduction in \texttt{CX} gate count and circuit depth relative to the original QPIE construction.
For all truncation levels $r<4$, none of the images satisfy either tolerance threshold, indicating a clear transition to visibly degraded reconstructions.

These results confirm that while JQPIE closely reproduces the behavior of classical JPEG at mild truncation levels, its reconstruction quality deteriorates sharply once the combined effects of coefficient quantization and truncation dominate.

In contrast to JQPIE, the bottom row of Fig.~\ref{fig:PSNR_SSIM_JQPIE_&_QF_misc} reveals a qualitatively different reconstruction behavior for QF-JQPIE.
By eliminating coefficient quantization and retaining only truncation of zigzag-ordered DCT coefficients, QF-JQPIE frequently achieves reconstruction quality that matches or exceeds the classical JPEG baseline.
At $r=5$, nearly all miscellaneous images exhibit PSNR values above the JPEG reference, with improvements reaching up to $10\,\mathrm{dB}$ in extreme cases.
A similar trend is observed for SSIM, where several reconstructions exceed the JPEG baseline by as much as $0.1$.

As truncation becomes more aggressive, QF-JQPIE maintains a markedly different degradation profile compared to JQPIE.
As shown in Fig.~\ref{fig:PSNR_SSIM_JQPIE_&_QF_misc} (bottom row), at $r=4$, $9$ images still achieve PSNR values at or above the JPEG baseline, while $25$ images remain above the SSIM reference level.
At $r=3$, no image satisfies the PSNR criterion; however, $2$ images continue to meet the SSIM tolerance.
This behavior can be attributed to the absence of quantization noise: truncation in an orthonormal transform domain effectively acts as a low-pass projection, preserving dominant low-frequency structures while suppressing high-frequency components that are less perceptually significant.
\begin{figure}[h]
	\centering
	\includegraphics[width=0.9\linewidth]{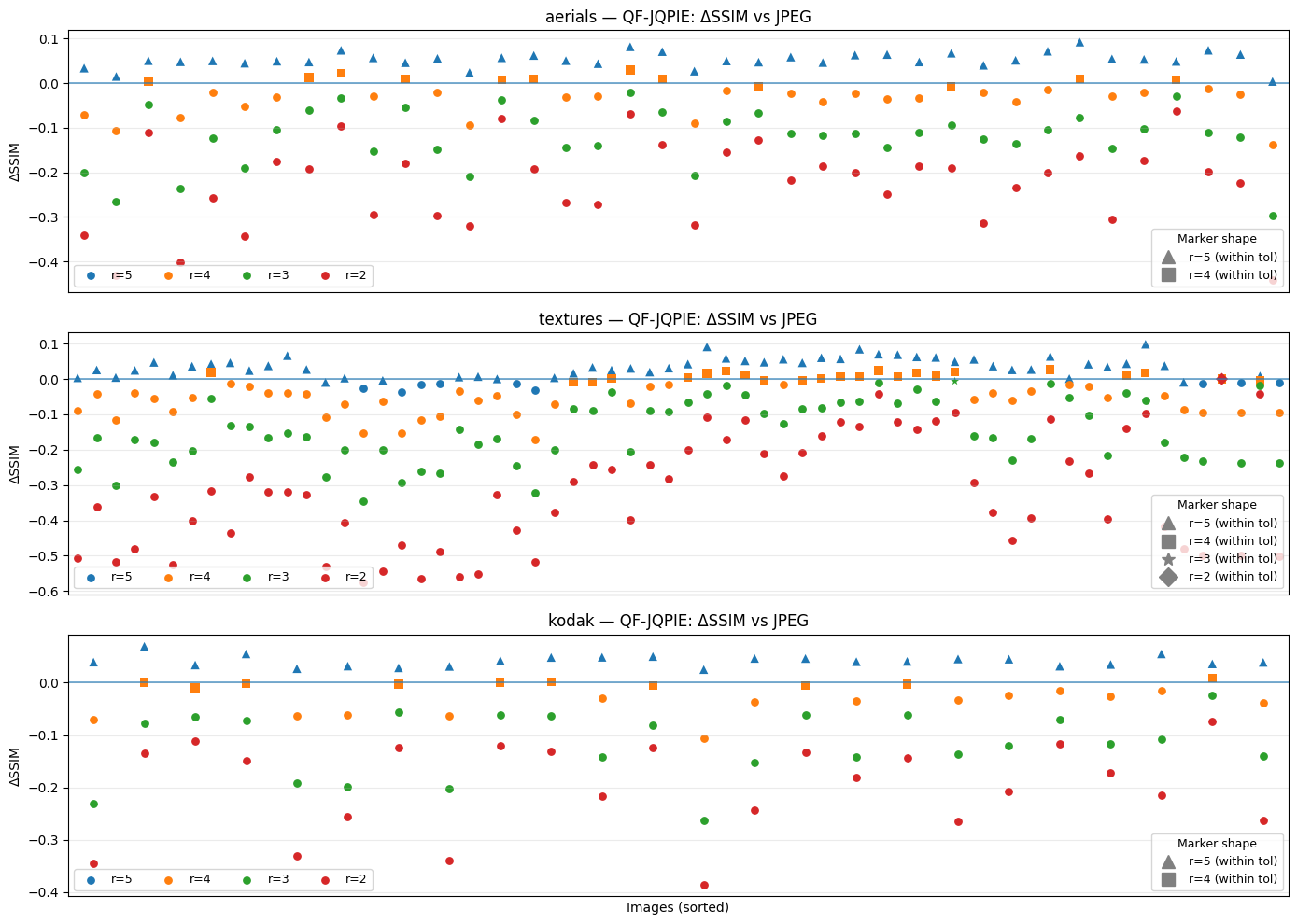}
	\caption{
		Structural similarity performance of the QF-JQPIE scheme measured by $\Delta$SSIM relative to classical JPEG decoding ($S=1$) for the aerial, texture, and Kodak image datasets.
		Each marker represents a single grayscale image, uniformly distributed along the horizontal axis.
		Positive deviations correspond to improved perceptual similarity compared to JPEG, demonstrating
		that truncation in the orthonormal DCT domain can preserve or enhance structural fidelity without
		quantization.
	}
	\label{fig:SSIM_JQPIE_&_QF_others}
\end{figure}

Further analysis across the aerial, texture, and Kodak image datasets confirms that the trends observed for the miscellaneous category are not dataset-specific.
Since JQPIE reconstructions consistently converge toward the JPEG baseline across these datasets, additional plots are omitted for brevity.
Instead, we focus on the QF-JQPIE results to demonstrate that this scheme can systematically achieve reconstruction quality comparable to or exceeding classical JPEG across diverse image classes.

Figures~\ref{fig:PSNR_JQPIE_&_QF_others} and~\ref{fig:SSIM_JQPIE_&_QF_others} show that, for certain texture images, aggressive truncation levels as low as $r=2$ remain perceptually comparable to JPEG.
At this truncation level, the \texttt{CX} gate count and circuit depth are reduced by approximately $96.75\%$ relative to the standard QPIE construction.
Such aggressive compression, however, is uncommon for natural images, as evidenced by the Kodak dataset, where acceptable reconstruction quality is achieved predominantly at $r=5$ for both hybrid schemes.

Beyond numerical reconstruction quality, the observed behavior highlights a fundamental difference between the two hybrid preparation strategies. Although both JQPIE and QF-JQPIE exploit DCT energy compaction and zigzag ordering to reduce quantum preparation cost, the role played by quantization is qualitatively different in a quantum amplitude-encoding setting.

In JQPIE, inverse quantization must be implemented using a block-encoded diagonal operator, which makes the procedure inherently probabilistic.
As a result, successful image reconstruction requires post-selection or OAA, increasing circuit depth and sampling overhead.
By contrast, QF-JQPIE avoids quantization entirely and relies only on truncation in an orthonormal transform domain.
Because amplitude encoding is sensitive to which basis states are populated rather than to the precise classical magnitudes of the coefficients, removing high-frequency components acts as a coherent low-pass filtering operation rather than as a source of distortion.
This explains why QF-JQPIE can both reduce circuit complexity and achieve PSNR and SSIM values that are comparable to, or higher than, those obtained with classical JPEG decoding.

From a complexity perspective, both JQPIE and QF-JQPIE exhibit the same asymptotic scaling for the dominant state-preparation stage, namely $\mathcal{O}(2^{2n-\ell})$, where $\ell$ denotes the number of inactive data-register qubits introduced by zigzag-domain truncation.
The distinction between the two schemes therefore does not arise from asymptotic gate complexity, but rather from constant factors and circuit structure.
In particular, QF-JQPIE eliminates the block-encoded inverse-quantization operator, avoids the use of an ancillary qubit, and removes the need for post-selection or OAA.

In this respect, the complexity of QF-JQPIE is comparable to other approximate image-loading strategies, such as the Fourier-series-based loader~\cite{moosa2023linear}, which achieves approximate image preparation with complexity $\mathcal{O}(2^{2m+4})$ for $m \le n-2$.
Crucially, QF-JQPIE attains similar asymptotic scaling while preserving the structured energy compaction of the DCT and exhibiting improved reconstruction fidelity in practice.
This combination of reduced circuit overhead, elimination of block-encoding-induced post-selection, and favorable reconstruction quality places QF-JQPIE in a qualitatively more advantageous operating regime than quantization-based hybrid approaches.

\section{Conclusion}\label{sec:conclusion}

In this work, we proposed a hybrid classical--quantum image preparation framework that integrates JPEG-style block compression with the QPIE amplitude-encoding representation, reducing the quantum resources required for preparing image states.
We developed two variants of the scheme.
The first, JQPIE, coherently implements the JPEG decompression workflow by realizing inverse quantization through a block-encoded diagonal operator.
The second, QF-JQPIE, removes quantization altogether and relies solely on zigzag-domain truncation, yielding a fully unitary and ancilla-free preparation procedure.

Statevector simulations on standard benchmark images (USC--SIPI and Kodak) show that, for practical truncation levels, both methods substantially reduce \texttt{CX} gate count and circuit depth relative to direct QPIE loading while maintaining high reconstruction quality measured by PSNR and SSIM.
Across all tested datasets, QF-JQPIE consistently achieves equal or better reconstruction quality than JQPIE, reflecting the fact that truncation in an orthonormal transform domain avoids the additional distortion and probabilistic overhead associated with block-encoded inverse quantization.

Overall, these results establish a concrete baseline for leveraging classical image transforms to make amplitude-based quantum image preparation more resource-efficient.
Future work includes extending the framework to alternative compression models such as JPEG2000 and wavelet-based transforms, and exploring data-driven representations (e.g., SVD/PCA and autoencoders) that may yield stronger coefficient concentration while preserving a reversible structure suitable for coherent quantum decompression.

\appendix
\section{Truncated implementation of the inverse zigzag permutation}\label{sec:appendix}

This appendix explains how the inverse zigzag permutation used in JPEG can be implemented in a truncated form that significantly reduces quantum gate count and circuit depth.  
To do this, within the JQPIE framework, only a limited number of low-frequency coefficients are retained after truncation. Thus 
it is unnecessary to implement the full dense zigzag permutation over all $64$ computational basis.

In standard JPEG compression, the zigzag transform is represented by a permutation matrix $P$ that reorders the vectorized $8\times8$ DCT coefficients according to the zigzag pattern shown in Fig.~\ref{fig:zigzag_reorder}.
In JQPIE, however, only the first $2^r$ coefficients in zigzag order are loaded into the quantum register.
This allows us to replace the full permutation $P$ with a truncated permutation $P_r$ that acts non-trivially only on the retained coefficients, while leaving all other basis states unchanged.

\subsection*{Example: truncation level $r=2$}

\begin{figure}[h]
	\centering
	\includegraphics[width=0.8\linewidth]{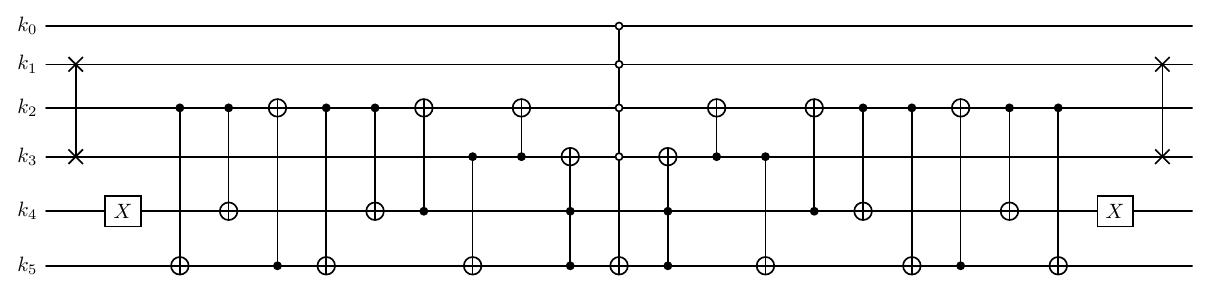}
	\caption{
		Quantum circuit implementing the truncated inverse zigzag permutation $P_2$ for truncation level $r=2$.
		The circuit permutes only the four retained low-frequency DCT coefficients while leaving all remaining basis states unchanged, thereby reducing gate count and circuit depth compared to the full zigzag permutation.
	}
	\label{fig:Permutation_2}
\end{figure}

We first illustrate the construction for $r=2$, where only four zigzag-ordered coefficients
are retained.
After zigzag reordering, the truncated coefficient vector for block $j$ is
\begin{equation}
	\hat{Z}_j^{(2)} =
	\big[
	\hat{C}_j(\pi(0)),
	\hat{C}_j(\pi(1)),
	\hat{C}_j(\pi(2)),
	\hat{C}_j(\pi(3))
	\big]
	=
	\big[
	\hat{C}_j(0),
	\hat{C}_j(1),
	\hat{C}_j(8),
	\hat{C}_j(16)
	\big].
\end{equation}
Accordingly, the inverse zigzag operation needs only to permute the indices $\{2,3,8,16\}$, while all remaining indices may be left fixed.
To ensure unitarity, the truncated permutation $P_2$ is defined as
\begin{equation}
	\pi(k)=
	\begin{cases}
		8,  & k=2,\\
		16, & k=3,\\
		2,  & k=8,\\
		3,  & k=16,\\
		k,  & \text{otherwise},
	\end{cases}
	\qquad k \in \{0,1,\ldots,63\}.
\end{equation}
This construction guarantees that $P_2$ remains a valid permutation matrix with exactly one non-zero entry per row and column, while avoiding unnecessary swaps involving inactive coefficients. 
The resulting quantum circuit implementing $P_2$ is synthesized using the Cybenko decomposition~\cite{cybenko2001reducing}. 
The quantum implementation of $P_2$ algorithm by the Qibo library~\cite{qibo_paper} has been depicted in Fig.~\ref{fig:Permutation_2}.

\begin{figure}[h]
	\centering
	\includegraphics[width=0.95\linewidth]{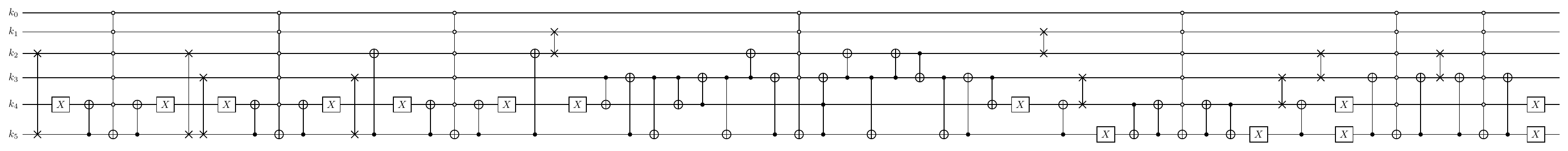}
	\caption{
		Quantum circuit realizing the truncated inverse zigzag permutation $P_3$ for truncation level $r=3$.
		Only the $2^3$ zigzag-ordered low-frequency coefficients are permuted, while inactive basis states are fixed, yielding a scalable and resource-efficient implementation consistent with truncated JPEG encoding.
	}
	\label{fig:Permutation_3}
\end{figure}

The same construction naturally extends to larger truncation parameters $r$. For each value of $r$, the truncated permutation $P_r$ is defined by identifying the $2^r$ zigzag indices that must be permuted and fixing all remaining basis states. Figure~\ref{fig:Permutation_3} illustrates the circuit implementation of $P_3$, corresponding to $r=3$.

In practice, the cost of implementing $P_r$ scales with the number of non-trivial swaps rather than with the full dimension of the permutation.
This leads to a substantial reduction in both gate count and circuit depth compared to the exact zigzag permutation, while remaining fully consistent with the truncated JPEG representation employed in JQPIE.

\section*{Funding}
This research did not receive any specific grant from funding agencies in the public, commercial, or not-for-profit sectors.



\bibliographystyle{cas-model2-names}

\bibliography{cas-refs}

\end{document}